\documentclass[11pt,a4paper]{article}
\usepackage{jcappub}
\usepackage{tikz}
\usepackage{epsfig}
\usepackage{amssymb}
\usepackage{subfigure}

\renewcommand\({\left(}
\renewcommand\){\right)}

\newcommand{\tr}{\text{tr}}
\newcommand{\diag}{\text{diag}}

\newcommand{\matrixx}[1]{\begin{pmatrix} #1 \end{pmatrix}} 

\newcommand{\gagamma}{g_{a\gamma}}

\newcommand{\J}[1]{\color{blue} #1 \color{black}}

\begin{document}


\begin{flushright}
{\large \tt DESY 17-116}
\end{flushright}


\title{
Stellar Recipes for Axion Hunters}

\author[a]{Maurizio Giannotti}
\author[b]{Igor G. Irastorza}
\author[b,c]{Javier Redondo}
\author[d]{Andreas~Ringwald}
\author[d]{Ken'ichi~Saikawa}

\affiliation[a]{Physical Sciences, Barry University,
11300 NE 2nd Ave., Miami Shores, FL 33161, USA}
\affiliation[b]{Departamento de F\'isica Te\'orica, Universidad de Zaragoza, Pedro Cerbuna 12, E-50009, Zaragoza, Espa\~na}
\affiliation[c]{Max-Planck-Institut f\"ur Physik (Werner-Heisenberg-Institut), F\"ohringer Ring 6, D-80805 M\"unchen, Germany}
\affiliation[d]{Theory Group, Deutsches Elektronen-Synchrotron DESY, Notkestra\ss{}e 85, D-22607 Hamburg, Germany}

\emailAdd{mgiannotti@barry.edu}
\emailAdd{igor.irastorza@cern.ch}
\emailAdd{jredondo@unizar.es}
\emailAdd{andreas.ringwald@desy.de}
\emailAdd{kenichi.saikawa@desy.de}

\abstract{
There are a number of observational hints from astrophysics which point to the existence of
stellar energy losses beyond the ones accounted for by neutrino emission. These excessive energy
losses may be explained by the existence of a
new sub-keV mass pseudoscalar Nambu--Goldstone boson with tiny couplings to photons, electrons, and nucleons.
An attractive possibility is to identify this particle
with the axion -- the hypothetical pseudo Nambu--Goldstone boson
predicted by the Peccei--Quinn solution to the strong CP problem.
We explore this possibility in terms of a DFSZ-type axion and of a KSVZ-type axion/majoron, respectively.
Both models allow a good global fit to the data,
prefering an axion mass around 10 meV.
We show that future axion experiments -- the fifth force experiment ARIADNE and the helioscope IAXO --
can attack the preferred mass range from the lower and higher end, respectively. An axion in this mass range
can also be the main constituent of dark matter.
}

\maketitle
\flushbottom

\section{Introduction}
\label{sec:introduction}
There is a strong physics case for the existence of the axion~\cite{Weinberg:1977ma,Wilczek:1977pj}. It
occurs in the arguably most elegant solution of the strong CP problem
as a pseudo Nambu-Goldstone
boson of a spontaneously broken global Abelian symmetry~\cite{Peccei:1977hh}.
The axion is also regarded as one of the best candidates of dark matter in the universe~\cite{Preskill:1982cy,Abbott:1982af,Dine:1982ah}.

The couplings and the mass of the axion are inversely proportional to the symmetry breaking scale.
Non-observation of axion signatures in laboratory experiments have pushed the latter much beyond the
electroweak scale, rendering the axion a very weakly interacting slim particle  (WISP)~\cite{Jaeckel:2010ni,Ringwald:2012hr}.
Such particles could be produced
in hot astrophysical plasmas, thus transporting energy out of
stars and other astrophysical objects.
In fact, the coupling strength of these particles with normal matter
and radiation is bounded by the constraint that
stellar lifetimes and energy-loss rates should not conflict with observations~\cite{Raffelt:1996wa}.
Such astrophysical limits have been derived from our Sun; from evolved low-mass stars, such as red giants and horizontal-branch stars in globular clusters, or
white dwarfs; from neutron stars; and from the duration of the neutrino burst of the core-collapse supernova SN1987A.

Intriguingly, there exist also marginal hints for additional energy losses in stars at different evolutionary stages -- red giants, supergiants, helium core burning stars, white dwarfs, and neutron stars~\cite{Ringwald:2015lqa,Giannotti:2015dwa,Giannotti:2015kwo}.
Indeed, the current observations show in all cases a preference for an additional cooling -- in most cases of the same order of magnitude as the standard one.

Throughout the years these results, globally known as the \textit{cooling anomaly problem}, have led to the speculation that the apparent systematic tendency of stars to cool faster than predicted originates from the production of new WISPs with tiny couplings to photons, electrons, and nucleons~\cite{Ringwald:2015lqa,Giannotti:2015dwa,Giannotti:2015kwo}.
Clearly, given the large systematics in stellar modeling and observations and the small statistical significance of the individual hints, any conclusion about new physics is certainly premature. However, with the current status of understanding, the WISP option cannot be ruled out and, in fact, would provide possibly the simplest solution to the cooling anomaly problem. Remarkably, according to the recent investigation in ref.~\cite{Giannotti:2015kwo}, axions and more generally axion-like particles (ALPs) would provide very good fits for the combined hints.

Most importantly, recent years have witnessed a revival in the experimental effort in axion/ALP  searches with an increasing interest in the development of a new generation of laboratory experiments. As a consequence, the parameter region hinted by the cooling anomalies is finally becoming accessible to  terrestrial experiments, such as the light-shining-through-walls experiment ALPS II \cite{Bahre:2013ywa}, the long-range force experiment ARIADNE~\cite{Arvanitaki:2014dfa}, and the helioscope  IAXO~\cite{Armengaud:2014gea}.
Unfortunately, the cooling anomalies mainly hint at an axion coupling to electrons, while the mentioned experiments are critically sensitive to the axion-photon and the axion-neutron coupling\footnote{Current and future weakly interacting massive particle (WIMP) direct detectection dark matter experiments,  such as XENON100~\cite{Aprile:2014eoa}  and
DARWIN~\cite{Aalbers:2016jon}, respectively, can also measure solar axions through electron ionisation signals induced by the axion-electron coupling. However, their sensitivities do not reach the level to explain the cooling hints.}.
In a concrete axion model, these couplings are related by a few constants, and thus we can study the detectability and complementarity of the experiments and obtain additional constraints in a model-dependent basis.
The aim of this paper is to select a sample of generic axion models and study whether they can accommodate all or a part of the stellar cooling  anomalies, providing the best fits of their parameters to guide their experimental discovery or refutation.

The paper is organized as follows. In section \ref{sec:cooling_summary} we summarize the observations of cooling anomalies and their interpretation in terms of model-dependent axion/ALP couplings to electrons, photon, neutrons, and protons. In section \ref{sec:axion_interpretation} we confront the required couplings with the predictions of  axion
models
and compare those with the capabilities of upcoming experimental searches.
In section \ref{sec:dm} we show that the axion hinted at by the anomalous stellar cooling may also
constitute the dark matter in the universe.
We summarize our results in section \ref{sec:conclusions}.
We leave the discussion of some technical details to the appendixes.
In appendix \ref{app:global}, we describe how we computed the $\chi^2$ for the global fits.
We present the state-of-the-art interpretation of the supernova (SN) bound on the nucleon couplings in appendix  \ref{app:sn1987a}.
The sensitivities of IAXO and ARIADNE are discussed in details in appendix \ref{app:Sensitivity}.

\section{Axion/ALP interpretation of stellar cooling anomaly observations}
\label{sec:cooling_summary}

Axions, and more generally ALPs, are pseudo Nambu-Goldstone bosons arising from the spontaneous breaking of an Abelian
global symmetry at a scale much larger than the electroweak scale.
At very low energies, their interactions with photons ($\gamma$), electrons ($e$), protons ($p$), and neutrons
($n$) can be
described by the Lagrangian
\begin{equation}
{\mathcal L} = \frac{1}{2} \partial_\mu a \partial^\mu a - \frac{1}{2} m_a^2 a^2
- \frac{\alpha}{8\pi} \,\frac{C_{a\gamma}}{f_a}\,a\,F_{\mu\nu} {\tilde F}^{\mu\nu}
+
\frac{1}{2}\, \frac{C_{af}}{f_a} \,
\partial_\mu a\ \overline{\psi}_f \gamma^\mu
\gamma_5   \psi_f
\label{axion_leff}
,
\end{equation}
where $f=e,p,n$; $F$ denotes the electromagnetic field strength tensor and $\tilde F$ its dual.
The model dependence rests in the mass $m_a$, the dimensionless coefficients $C_{ai}$, and the decay constant $f_a$, which is proportional to the symmetry breaking scale.

Hints for excessive cooling have been observed in: \textit{1)} several pulsating white dwarfs (WDs), whose cooling efficiency was extracted from the rate of the period change;
\textit{2)} the WD luminosity function (WDLF), which describes the WD distribution as a function of their brightness;
\textit{3)} red giants branch (RGB) stars in globular clusters, in particular the luminosity of the tip of the branch;
\textit{4)} horizontal branch (HB) stars in globular clusters or, more precisely, the $R$-parameter, that is the ratio of the number of HB over RGB stars;
\textit{5)} helium burning supergiants in open clusters, more specifically the ratio $B/R$ of blue and red supergiants;
and \textit{6)} neutron stars (NS).
Table~\ref{tab:anomalies} lists the results and their references.

\begin{table}[t]
	\begin{center}
		\begin{tabular}{ l  c  l   l}
			\hline \hline
			\textbf{Observable} &	& \textbf{Stellar System}					& \textbf{References}	\\ \hline
			Rate of	period	change	&	& WD Variables  						& \cite{Corsico:2012ki,Corsico:2012sh,Corsico:2014mpa,Corsico:2016okh,Battich:2016htm}	\\
 \hline
			Shape of WDLF		&	& WDs					   					&\cite{Bertolami:2014noa,Bertolami:2014wua}\\ \hline
			Luminosity of 	the	RGB tip	&	& Globular Clusters		   					&\cite{Viaux:2013hca,Viaux:2013lha,Arceo-Diaz:2015pva}	\\
 \hline
			$R$-parameter			&	& Globular Clusters  	 	 	 			&\cite{Ayala:2014pea,Straniero:2015nvc}\\ \hline
			$B/R$					&	& Open Clusters      						&\cite{Skillman:2002aa,McQuinn:2011bb,Friedland:2012hj,Carosi:2013rla}\\ \hline
			NS surface temperature evolution		&	& Cassiopeia A				 	  					&\cite{Leinson:2014ioa}\\ \hline \hline
		\end{tabular}
		\caption{Summary of anomalous cooling observations (from~\cite{Giannotti:2016hnk}).
			}
		\label{tab:anomalies}
	\end{center}
\end{table}

The most relevant axion/ALP production mechanism for stars with a high density core, such as WDs and RGB stars, is bremsstrahlung off electrons,
\begin{equation}
\label{Eq:ALP_bremm}
e+Ze\to Ze+e+a\,,
\end{equation}
which induces an additional energy loss rate proportional to $g_{ae}^2 T^4 $, where
\begin{equation}
g_{af}^2\equiv C_{af}^2\,m_f^2 /f_a^2; \hspace{3ex} f=e,p,n .
\end{equation}

As shown in ref.~\cite{Giannotti:2015kwo}, this temperature dependence is optimal to fit the WDLF and provides a reasonably good explanation for the observed excess cooling in DA and DB WD variables, whose internal temperatures differ by a factor of a few.
The combined analysis of all the observed WD variables gives a fairly good fit,
$\chi^2_{\rm min}/ $d.o.f.$ =1.1 $, for $ g_{ae}=2.9\times 10^{-13} $ and favours the axion/ALP interpretation at $2\,\sigma $.
Moreover, the peculiar temperature dependence of the axion bremsstrahlung rate allows to account  for the excessive cooling observed in RGB stars~\cite{Viaux:2013lha}, which have a considerably larger internal temperature than WDs, with a comparable axion-electron coupling.
For this paper we have reexamined the WDLF (see appendix \ref{app:global}) and combined it with the WD pulsation, and RGB stars.
Our fit gives
\begin{equation}
g_{ae}=1.6^{+0.29}_{-0.34} \times 10^{-13},
\end{equation}
with $ \chi^2_{\rm min}/ $d.o.f.$ =14.9/15=1.0$, and favours the axion/ALP solution at slightly more than $3\,\sigma $.

Axion bremsstrahlung off electrons contributes also to the evolution of low mass stars~\cite{Giannotti:2015kwo}.
Its main effect is to delay helium ignition at the end of the RGB phase while leaving essentially unchanged the HB stage, thereby reducing the expected ratio of HB and RGB stars (the $R$-parameter).
The most recent study of a set of globular clusters outputs a strong constraint, but also a mild preference for additional cooling~\cite{Ayala:2014pea,Straniero:2015nvc}.

The reduction of $R$ could also be due to other axion emission channels.
Most notably, the Primakoff process, consisting in the conversion of a photon into an axion/ALP in the electric field of nuclei
and electrons,
\begin{eqnarray}
\gamma +Ze\to Ze+a\,,
\end{eqnarray}
is the most discussed one in relation to the hint.
This process depends strongly on the environment temperature and is suppressed at high density (e.g., those characterizing the core of WDs and RGB stars) by the plasma frequency and degeneracy effects (see, e.g., \cite{Raffelt:1987yu}).
However, it could efficiently accelerate the cooling of the HB stars and consequently lower the observed $R$-parameter.

\begin{figure}[t]
\centering
\includegraphics[width=0.5\textwidth]{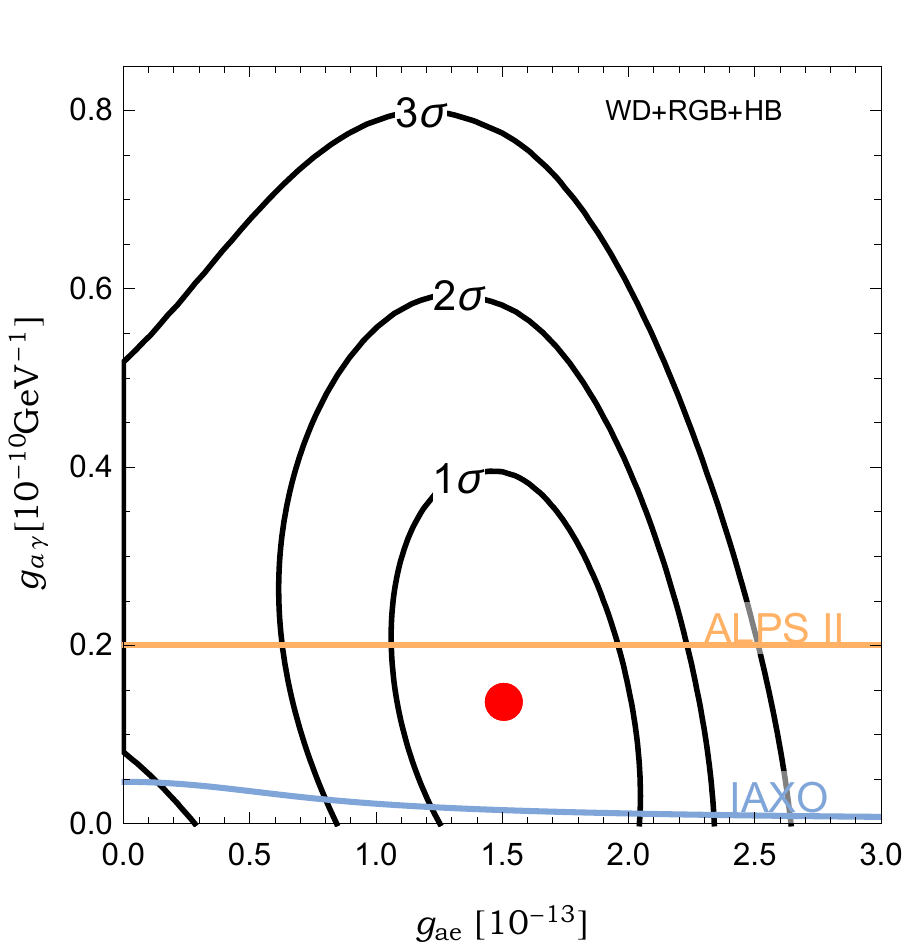}
\caption{Combined analysis of the hints from WD+RGB+HB stars in the $ g_{ae} - g_{a\gamma} $ plane.
Also shown are the projected sensitivities of the light-shining-through-walls experiment ALPS II \cite{Bahre:2013ywa} and the
helioscope IAXO~\cite{Armengaud:2014gea}.
}
\label{fig:hinted_region}
\end{figure}

In figure~\ref{fig:hinted_region}, we display  the 1, 2, and $3\,\sigma$ hinted areas for the combined fit of the
WD, RGB, and HB cooling anomalies\footnote{Notice that the constraints (and hints) from stellar evolution that we are reporting here are calculated in the approximation of vanishing axion mass. In fact, the axion production rate in a star does not depend on the axion mass unless $ m_a\gg T$, where $ T $ is the stellar core temperature. In this case, the axion production in the stellar core would be Boltzmann suppressed. Since the coldest stars we are considering here are WDs with a core temperature of $ \sim  $ keV, we can assume that our results are valid up to $ m_a\sim $ a few keV.} in the
\begin{equation}
g_{a\gamma}\equiv \frac{\alpha}{2\pi f_a} |C_{a\gamma}|
\end{equation}
vs. $g_{ae}$ plane.
Note, that the quantitative analysis with the data at hand shows that it is possible to explain all the observed cooling hints, including the $R$-parameter anomaly, even neglecting the axion-photon coupling, though there is a preference for non-vanishing couplings with both electrons and photons.
The best fit value, $g_{ae}=1.5\times 10^{-13} $ and $ g_{a\gamma}=0.14\times 10^{-10} $ GeV$ ^{-1} $,
is indicated in the figure with a red dot.
Taken at face value, it implies $|C_{ae}/C_{a\gamma}|=0.025$, which is a somewhat unusual ratio because typical axion models have
either $|C_{ae}/C_{a\gamma}|\sim {\cal O}(1)$ or loop-suppressed, $\sim (3\alpha^2/2\pi)\log (f_a/m_e)\sim {\cal O}(10^{-3})$.
We will explore axion models of both these typical types and consider also axion-majoron models
where the above unusual ratio can be obtained. Note, however, that the 1\,$\sigma$ contour is compatible with a zero axion-photon coupling.

An additional hint comes from the anomalously fast decline of the surface temperature of the
neutron star in the supernova remnant Cassiopeia A (CAS A), which can be interpreted as due to
axion emission in neutron ${}^3$P$_2(m_j=0)$ Cooper pair formation
with a neutron coupling~\cite{Leinson:2014ioa}
\begin{equation}
g_{an}^2  \sim (1.4\pm0.5)\times 10^{-19}.
\end{equation}
The estimate of~\cite{Leinson:2014ioa} does not come with an error bar, though the study claims that values outside the quoted range cannot fit the data. We shall warn that the systematics in this calculation are very hard to estimate. Indeed, the cooling of the superfluid core in the neutron star may be also explained by other means, for instance, by neutrino emission in pair formation in a multicomponent superfluid state ${}^3$P$_2(m_j=0,\pm 1, \pm 2)$~\cite{Leinson:2014cja}.
Thus, at the moment, we cannot take seriously this NS hint, and we can only ask the question whether the hinted parameters can be compatible with the other hinted regions in particular models.
To this end, we will confront axion models with stellar cooling hints with and without including the NS hint. For these purposes it will suffice to interpret the quoted range as a 1\,$\sigma$ interval.
Note that the range quoted is compatible with the upper limit $g_{an}  <  8\times 10^{-10}$ from NS cooling in refs. \cite{Keller:2012yr,Sedrakian:2015krq}.
This, however, is more complicated to interpret in terms of a combination of couplings.
In fact, it includes also axion emission in nucleon bremsstrahlung $N+N\to N+N+a$ where $N$ can be either a proton or a neutron and most of the simulations were done with a very small neutron coupling, so that the effect is mostly due to the proton coupling.

Finally, the axion/ALP bremsstrahlung off nucleons can shorten the prediction of the neutrino pulse duration of core collapse supernovae. In fact, the neutrino observations from SN1987A lead to a bound (see appendix \ref{app:sn1987a})
\begin{align}
\label{nu_pulse_sn1987a}
g_{ap}^2+g_{an}^2 < 3.6\times 10^{-19}\,.
\end{align}
We will consider this as a $1\,\sigma $ hint that $ g_{aN}^2=0 $ within the error $3.6\times 10^{-19} $.
However, we warn the reader that SN 1987A constraints are based on axion emissivities not completely understood and on simulations that at the moment do not include all necessary physics and therefore have systematic uncertainties themselves. Again, we will study axion models with and without including this constraint.

\section{Axion interpretation of stellar cooling anomalies}
\label{sec:axion_interpretation}

Let us start by reviewing the generic features of the axion.
The basic building block of an invisible axion model is a global $U(1)_{\rm PQ}$ symmetry, which
is broken at a high scale by the vacuum expectation value $\langle\sigma^2\rangle =v_{\rm PQ}^2/2$
of a complex Standard Model (SM) singlet scalar field $\sigma$.
In this notation, the axion field appears as the phase of this complex scalar $\sigma=(v_{\rm PQ}/\sqrt{2})e^{i a/v_{\rm PQ}}$
or as a linear combination of this and other Higgs phases.
The associated Noether current  $J_\mu^{\rm PQ}$ is required to have
a color anomaly and, although not required for solving the strong CP
problem,  it may also have an electromagnetic anomaly:
\begin{equation}
\label{eq:NE}
\partial^\mu J_\mu^{PQ} =
\frac{N \alpha_s}{8\pi} G^a_{\mu\nu} \tilde G^{a\mu\nu} +
\frac{E \alpha}{8\pi} F_{\mu\nu} \tilde F^{\mu\nu} \,,
\end{equation}
where $G^a_{\mu\nu}$ is the color field strength
tensor, $\tilde G^{a\mu\nu}$ its dual, while
$N$ and $E$ are anomaly coefficients.
The decay constant of the associated
Nambu-Goldstone boson is then given by
\begin{equation}
f_a = \frac{v_{\rm PQ}}{N}.
\end{equation}
The axion mass $m_a$, in units of  the decay constant $f_a$, is equal to  the square root of the topological
susceptibility $\chi$ in QCD. Recent precision calculations of the latter by next-to-leading order
chiral perturbation theory~\cite{diCortona:2015ldu}  or lattice
simulations~\cite{Borsanyi:2016ksw}
result in
\begin{equation}
\label{zeroTma}
m_a\equiv \frac{\sqrt{\chi}}{f_a} =
{5.70(7)\,   \left(\frac{10^{9}\,\rm GeV}{f_a}\right)\text{meV}. }
\end{equation}
Moreover, the photon coupling of the axion is given by
\begin{equation}
C_{a\gamma}= \dfrac{E}{N}-1.92(4)\,,
\label{eq:photon_coupling}
\end{equation}
while the proton and neutron have a model-independent part and
a model dependent contribution that arises from possible axion-quark couplings
of the form $(C_{aq}/2)(\partial_\mu a/f_a)\bar \psi_q \gamma^\mu\gamma_5\psi_q$
in the high-energy theory,
\begin{align}
\label{eq:nucleon_couplings}
C_{ap} & = -0.47(3)+0.88(3) C_{au} -0.39(2) C_{ad}-0.038(5)C_{as} \nonumber \\
				&\hspace{2.15cm}-0.012(5) C_{ac} -0.009(2) C_{ab}-0.0035(4) C_{at}\,, \nonumber  \\
C_{an} & = -0.02(3)+0.88(3) C_{ad} -0.39(2) C_{au}-0.038(5)C_{as} \nonumber \\
				&\hspace{2.15cm}-0.012(5) C_{ac} -0.009(2) C_{ab}-0.0035(4) C_{at} \,,
\end{align}
as found in the state-of-the-art calculation~\cite{diCortona:2015ldu}.

Note that the couplings $g_{ai}\propto 1/f_a$ of the axion to $i=\gamma,e,p,n$ are effectively proportional to its mass,
 $g_{ai}\propto m_a$.
That is why, in specific axion models, the coupling strength is often parameterized with the mass\footnote{We follow this approach in our figures~\ref{fig:hinted_region_DFSZ_tanbeta_m} and \ref{fig:hinted_region_KSVZ_AJ_all}, where we show the mass scale on the $x$-axis.
 		The mass there emerges solely from its relation with $ f_a $, Eq.~\eqref{zeroTma}. We remind, however, that the stellar hints are calculated in the approximation of masseless axions, as explained in footnote 2.}.
ALPs generically do not feature this particular relation between their couplings and their mass.

\subsection{DFSZ axion}
\label{sec:dfsz}

In the DFSZ axion model~\cite{Zhitnitsky:1980tq,Dine:1981rt}, the Standard Model (SM) Higgs sector is extended
to contain two Higgs doublets, $H_u$ and $H_d$, whose vacuum expectation values $v_u$ and $v_d$ give masses to up-type
and down-type quarks, respectively.
There are two possibilities,  dubbed DFSZ I or DFSZ II, according to whether leptons couple to $H_d$, which occurs in familiar Grand Unified Theories (GUT), or to $H_u$. In the usual nomenclature \cite{Olive:2016xmw}, the Higgs sector of DFSZ I is a
type-II Two Higgs Doublet Model (2HDM), which is easily embedded in a GUT, while the one of DFSZ II is a type-IV, sometimes also called Flipped 2HDM. Correspondingly, the
Yukawa interaction terms in DFSZ I read
\begin{align}
\label{lyuk_dfsz_I}
{\mathcal L_Y} = \Gamma_{ij}\overline{q}_{iL} H_d d_{j R}
+ Y_{ij}\overline{q}_{iL}\widetilde{H}_u u_{j R}
+ G_{ij}\overline{L}_{i } H_d l_{j R} + h.c.\,,
\end{align}
while, in DFSZ II,
\begin{align}
\label{lyuk_dfsz_II}
{\mathcal L_Y} = \Gamma_{ij}\overline{q}_{iL} H_d d_{j R}
+ Y_{ij}\overline{q}_{iL}\widetilde{H}_u u_{j R}
+ G_{ij}\overline{L}_{i } H_u l_{j R} + h.c.\,.
\end{align}
Here $\widetilde{H}_u=\epsilon H_u^*$,  $i,j=1,2,3$ are flavor indices and
$\Gamma_{ij}$, $Y_{ij}$,  $G_{ij}$ are complex
$3\times 3$ matrices.

\begin{figure}[t]
\centering
\vspace{-0.7cm}

\includegraphics[width=0.5\textwidth]{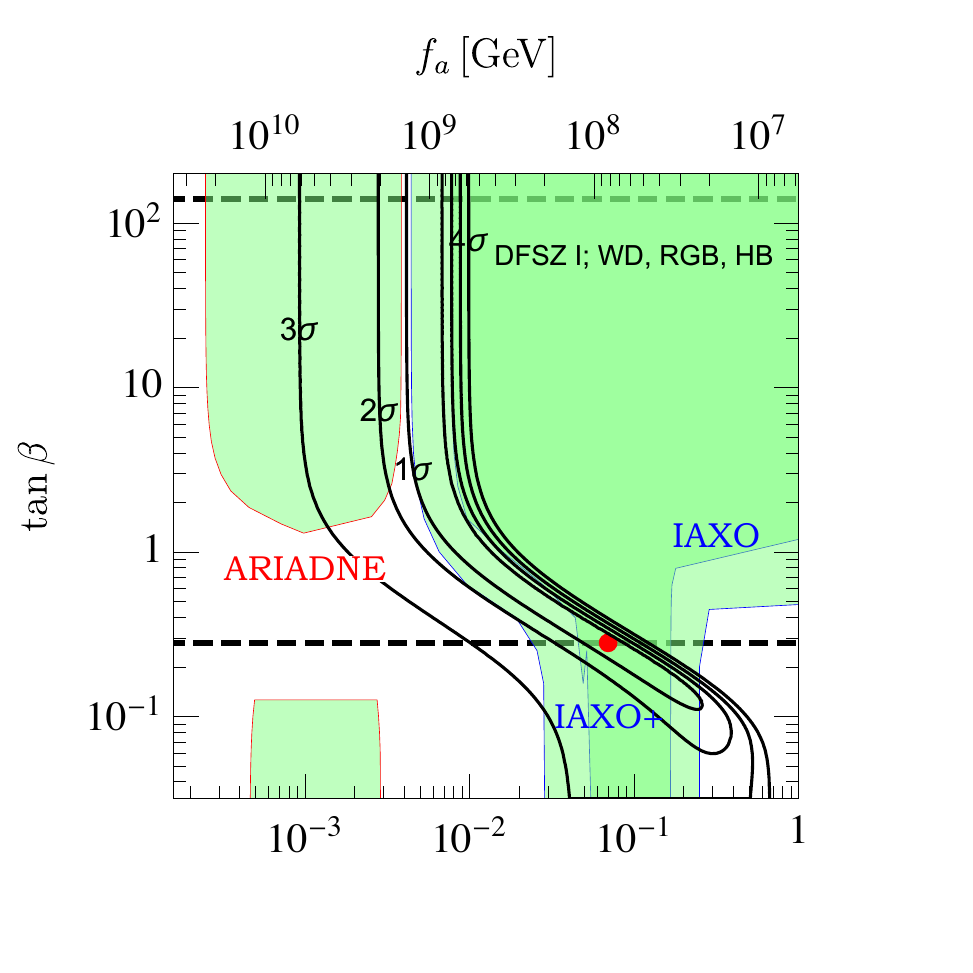}
\hspace{-1.8cm}
\includegraphics[width=0.5\textwidth]{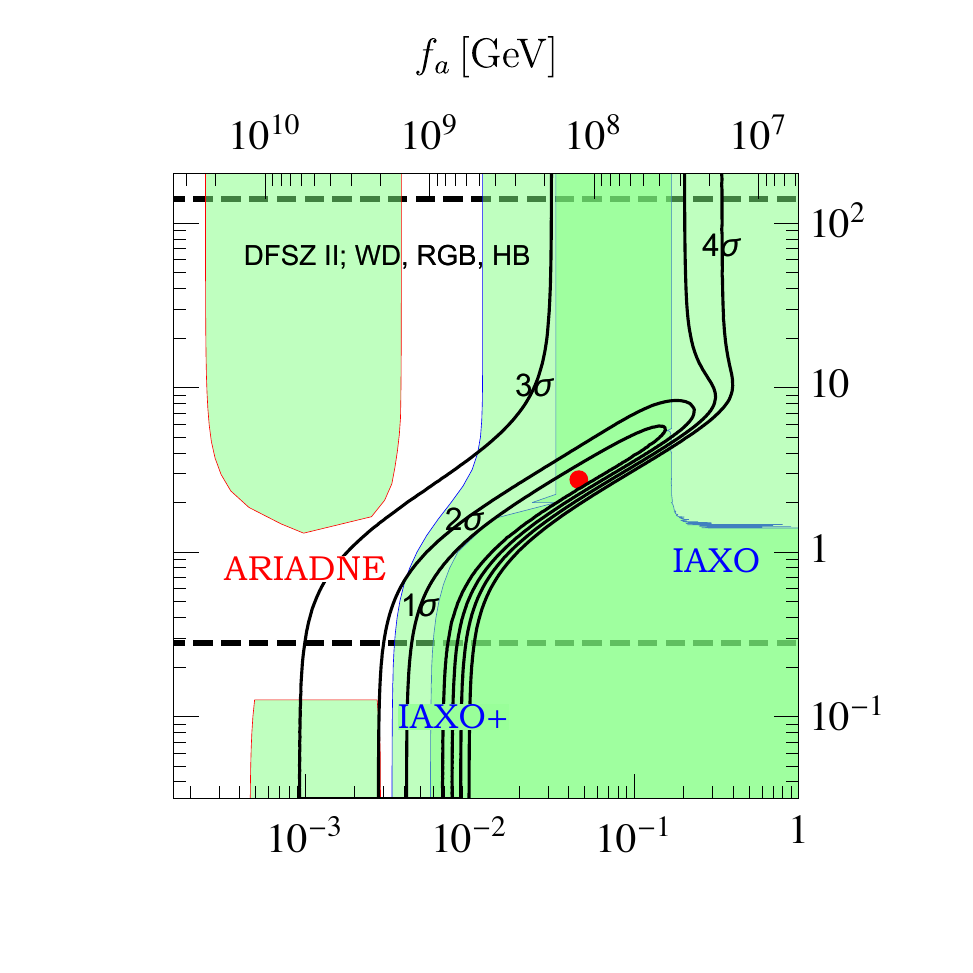}
\vspace{-1.4cm}

\includegraphics[width=0.5\textwidth]{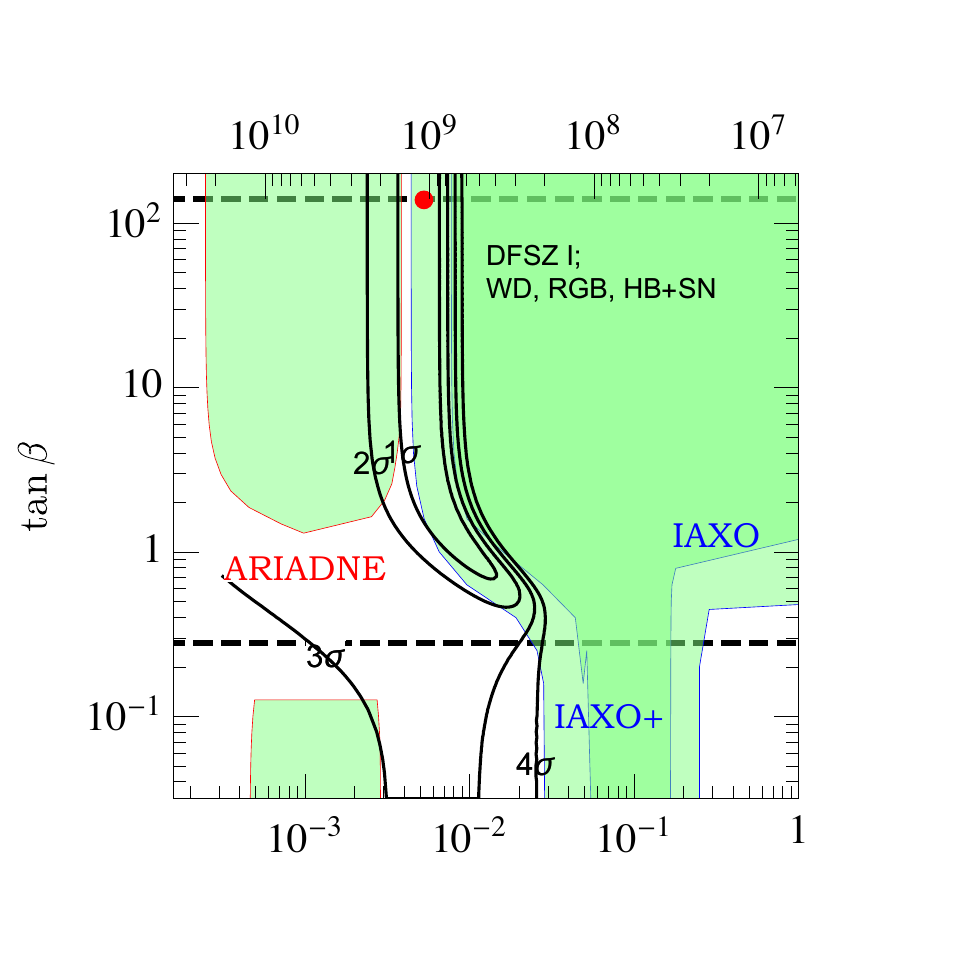}
\hspace{-1.8cm}
\includegraphics[width=0.5\textwidth]{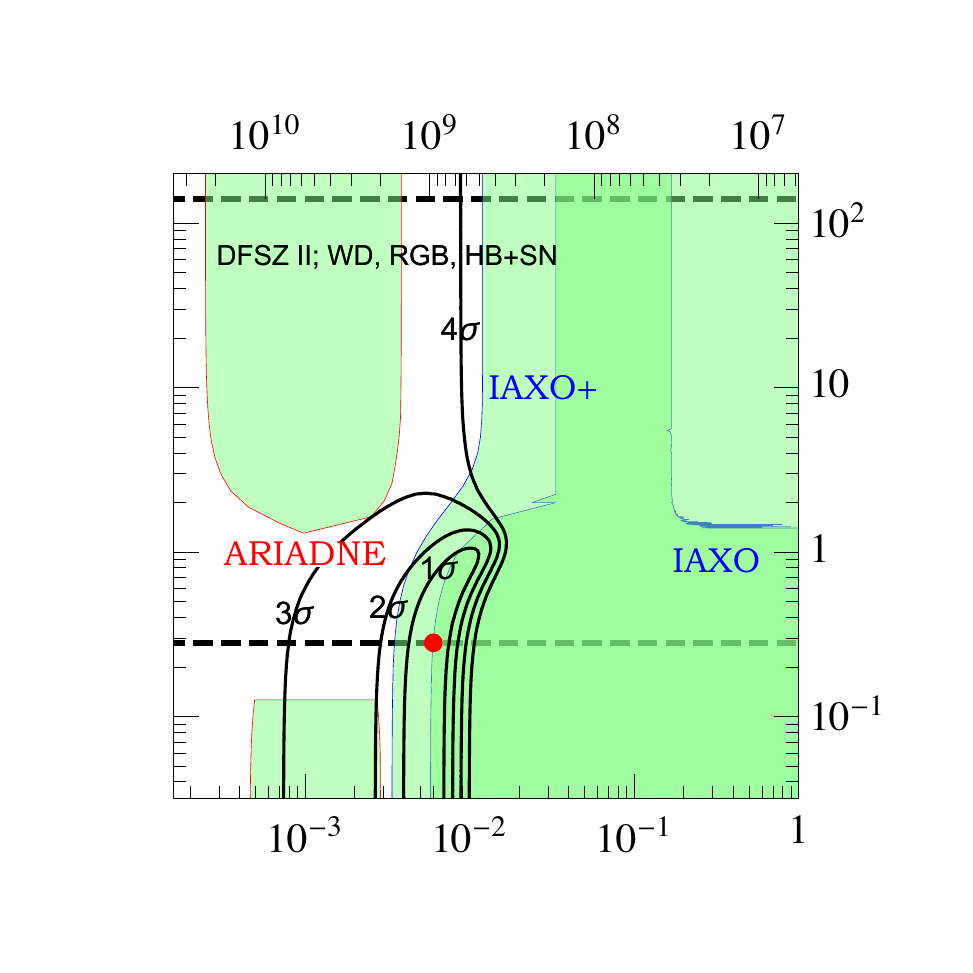}
\vspace{-1.4cm}

\includegraphics[width=0.5\textwidth]{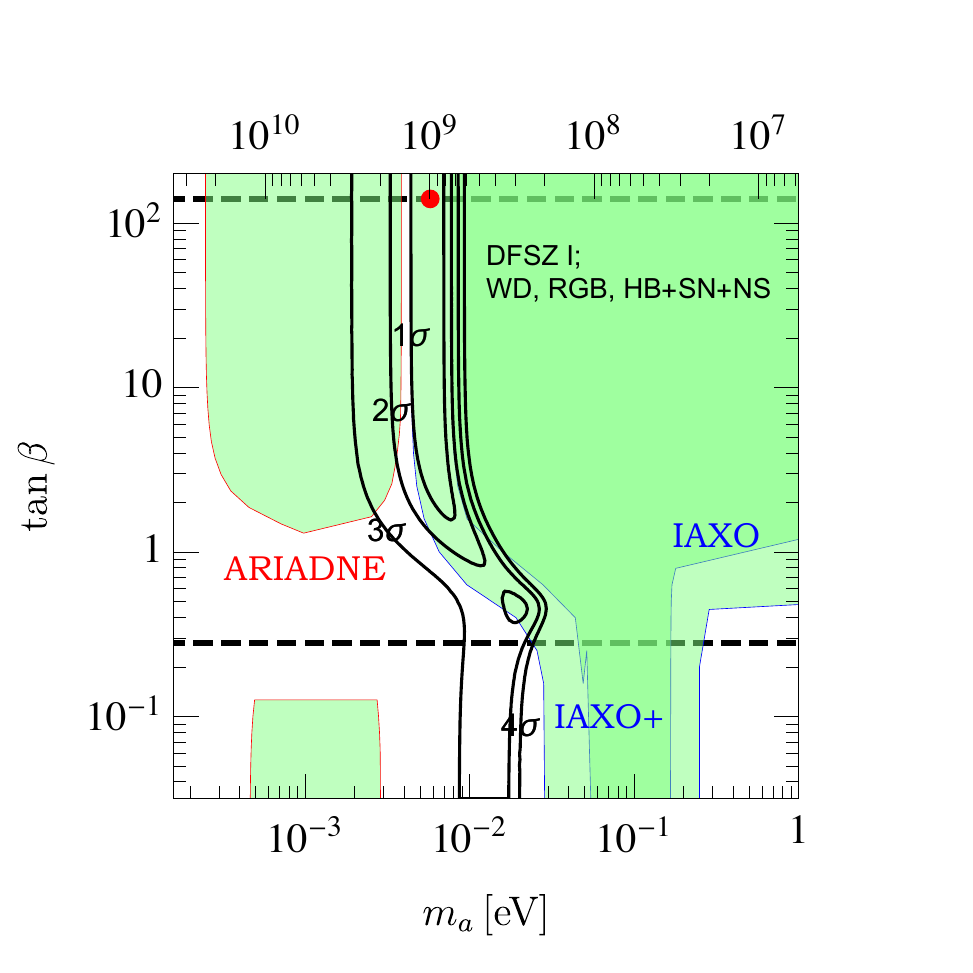}
\hspace{-1.8cm}
\includegraphics[width=0.5\textwidth]{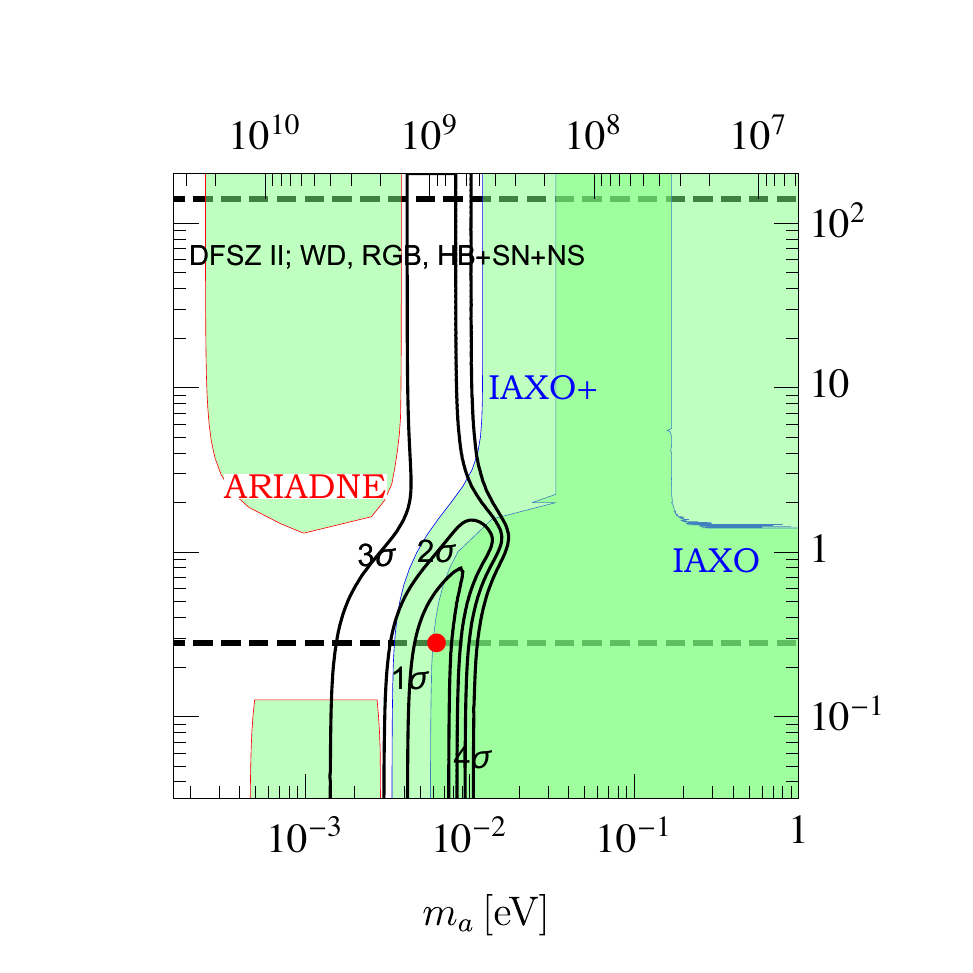}
\vspace{-.5cm}
\caption{
1, 2, 3, 4\,$\sigma$ contours in the global fit of DFSZ I (left panels) and II (right panels) with WD+RGB+HB data alone (top) and including the SN 1987A constraint (middle) and, in addition, the NS CAS A data (bottom).
Perturbative unitarity in the Yukawa couplings to fermions is satisfied for values of  $\tan\beta$  between the two dashed lines.
The red bullets are the best fits compatible with perturbative unitarity.
Also shown are the projected sensitivities of  ARIADNE \cite{Arvanitaki:2014dfa} and IAXO~\cite{Armengaud:2014gea}.}
\label{fig:hinted_region_DFSZ_tanbeta_m}
\end{figure}

The interactions given by eq.~\eqref{lyuk_dfsz_I} (DFSZ I)
 or eq.~\eqref{lyuk_dfsz_II}  (DFSZ II)
are assumed to be invariant under a U(1)$_{\rm PQ}$ symmetry with symmetry breaking scale $v_{\rm PQ}$.
At low energies, the effective Lagrangian is then given by eq.~\eqref{axion_leff}, with~\cite{Srednicki:1985xd,Dias:2014osa}
\begin{equation}
f_a = \frac{v_{\rm PQ}}{6},
\end{equation}
\begin{eqnarray}
&& C^{\rm DFSZ\ I}_{ae}= \frac13 \sin^2\beta\,, \qquad C^{\rm DFSZ\ II}_{ae}= \frac13 (1-\sin^2\beta)\,,  \\
\end{eqnarray}
and~\cite{diCortona:2015ldu}
\begin{eqnarray}
&& C^{\rm DFSZ\ I}_{a\gamma}= \dfrac{8}{3}-1.92(4) \,, \qquad  C^{\rm DFSZ\ II}_{a\gamma}= \dfrac{2}{3}-1.92(4) \,, \\
&& C_{Ap}=-0.435\sin^2\beta+\left(-0.182\pm 0.025\right)\,, \nonumber \\
&& C_{An}=0.414\sin^2\beta+\left(-0.16\pm 0.025\right)\,.
\end{eqnarray}
Here, $\tan\beta \equiv v_u/v_d$, with $v = \sqrt{v_u^2 + v_d^2}=246\ {\rm GeV}$. It is theoretically
 constrained from both ends by the requirement of perturbative unitarity of the Yukawa couplings,
\begin{equation}
0.28 < \tan\beta < 140\, .
\end{equation}
Here, the lower limit arises in all 2HDMs, while the upper limit is specific to the type-II and type-IV 2HDMs \cite{Chen:2013kt}.

The DFSZ models have only two parameters, $f_a$ and $\tan\beta$, that we can extract from the global fit of the WDLF, the period decrease of 4 pulsating WDs (R548, L 19-2	(113), L 19-2 (192), and PG 1351+489), the luminosity of the tip in the RGB of M5 and the $R$-parameter in globular clusters, which we hereafter label as HB, see appendix \ref{app:global} for specifics. The best fit values are recorded in table~\ref{tab:DFSZ_hints} and the
1, 2, 3, 4\,$\sigma$ contours are shown in figure~\ref{fig:hinted_region_DFSZ_tanbeta_m}.
Note that we impose the constraint on perturbative unitarity on the best fit values but not
on the contours.
The resulting regions can be understood as follows.
\begin{table}[h]
	\begin{center}
		\begin{tabular}{| l |l || c |  c| c | c|}
			\hline
			\textbf{Model} & Global fit includes	
							&  $f_a\, [10^8\,$GeV]
										& $ m_a  $ [meV]
												& $\tan\beta$ 	& $\chi^2_{\rm min}/{\rm d.o.f.}$	\\ \hline \hline
		& WD,RGB,HB 			&  0.77 		&  74   	& 0.28 		&  14.9/15				\\
DFSZ I 	& WD,RGB,HB,SN 		&  11		&  5.3  	& 140  		& 16.3/16 				\\
		& WD,RGB,HB,SN,NS 	&  9.9 		&  5.8  	& 140 		& 19.2/17 				\\ \hline
		& WD,RGB,HB 			&  1.2  		&  46 	& 2.7 		& 14.9/15  			\\
DFSZ II	& WD,RGB,HB,SN 		&  9.5  		&  6.0  	& 0.28 		& 15.3/16 				\\
		& WD,RGB,HB,SN,NS	&  9.1  		&  6.3  	& 0.28 		& 21.3/17				\\		\hline
		\end{tabular}
		\caption{Best fit parameters compatible with  perturbative unitarity for DFSZ-type axion interpretations of the cooling anomalies.
}
		\label{tab:DFSZ_hints}
	\end{center}
\end{table}
%

Let us consider first the DFSZ I case.
In figure~\ref{fig:hinted_region_DFSZ_tanbeta_m} (up, left) we show the 1, 2,  3, 4\,$\sigma$ contours
for the hints involving only the coupling to electrons and photons (WD+RGB+HB stars).
The main driver of the fit is the WDLF, forcing a concrete value for $g_{ae}=m_e\sin^2\beta /3f_a\sim 1.5\times 10^{-13}$ which
shows as a vertical line for large $\tan\beta$ ($\sin^2\beta\sim1$) with a diagonal turn towards the bottom right when the small $\tan\beta$ is compensated by a decrease of $f_a$.
The degeneracy is broken at $f_a\sim 3\times 10^7$ GeV because of the $g_{a\gamma}$ dependency of the $R$-parameter.
A small value of $f_a$ makes $g_{a\gamma}$  large and gives too small a value for the $R$-parameter.
The fact that the 1 $\sigma$ contour shows both the vertical and diagonal branches means that we can have large $\tan\beta$
and small $\tan\beta$ DFSZ I axions fitting the data.

However, adding the bound from SN 1987A forces the axion decay constant to be large, disfavouring the small $\tan\beta$ solution, $\tan\beta>0.7$ at 1 $\sigma$ (the best fit coincides with the unitarity constraint $\tan\beta\sim 140$),
and the regions become almost vertical, see ~\ref{fig:hinted_region_DFSZ_tanbeta_m} (medium, left).
The WD+RGB+HB set is in tension with the SN constraint and that worsens a bit the
$\chi^2_{\rm min}/{\rm d.o.f.}$.
In this large $\tan\beta$ region the axion-neutron coupling required to fit the NS hint from CAS A is at odds with the SN bound  ($C_{ap}$ is comparable but slightly larger than $C_{an}$) so the fit worsens by including it, see table \ref{tab:DFSZ_hints} and figure~\ref{fig:hinted_region_DFSZ_tanbeta_m} (bottom, left).
However, there is a second local minimum of $\chi^2$ with $\chi^2_{\rm min}/$d.o.f.$=24.8/17\sim 1.4$
at $\tan\beta\sim 0.5$ and $m_a\sim 20$ meV. At small $\tan\beta$ the SN constraint becomes weaker compared to the NS hinted region because the coupling to protons becomes of the same order as the coupling to neutrons. Of course, this solution makes sense only if we trust the NS hint, and somehow disregard or diminish the importance of the SN constraint.

Let us now discuss the DFSZ II case. The favoured region by WD+RGB+HB is still dominating the fit to give
$g_{ae}\sim 1.5\times 10^{-13}$. But now $g_{ae}=m_e(1-\sin^2\beta) /3f_a$, so small $\tan\beta$ corresponds to
a vertical line at the largest $f_a$ and large $\tan\beta$ can be compensated by a small $f_a$, corresponding to the
up-right diagonal branch in figure~\ref{fig:hinted_region_DFSZ_tanbeta_m} (up, right).
The SN constraint enforces again large $f_a$ so that now the small $\tan\beta$ solution is preferred, giving a better fit
than in DFSZ I because $C_{ap},C_{an}$ are relatively smaller at small $\tan\beta$, see table \ref{tab:DFSZ_hints}.
As in DFSZ I, the SN constraint cancels the hint for cooling in HBs due to the axion-photon coupling and thus the best fit region is the one further away from the SN constraint, so the best fit turns out to be in the unitarity constraint $\tan\beta\sim 0.28$, although the $1\sigma$ region extends to $\tan\beta<1$.
This fit is incompatible with the NS hint at more than $2\sigma$, so the fit degrades by including it, see table \ref{tab:DFSZ_hints}. The reason is not the tension with SN like in DFSZ I, but with the WD+RGB+HB set. In DFSZ I with large $\tan\beta$ we have $C_{ae}/C_{an}=(1/3)/(0.414-0.16)=1.3$ while in DFSZ II at small $\tan\beta$ the ratio is
$C_{ae}/C_{an}=(1/3)/(-0.16)=-2$. Thus, for the same electron coupling the neutron coupling is a factor $\sim 2$ weaker, which adds up to the already not good NS fit of DSFZ I at large $\tan\beta$.

On a pessimistic note, $f_a=\infty$, which corresponds to axion decoupling, is still allowed at $\sim 3-4$ $\sigma$, which is not very significant when we take into account that there might be unidentified systematics in the cooling analysis.
However, at 3 $\sigma$ the best fit regions indicate always a non-zero values of $f_a$ and some axion cooling at work.
We have shown the best fit as red points in figure~\ref{fig:hinted_region_DFSZ_tanbeta_m}, but we warn once more that they have little meaning.
The best guidance is provided by the 1 $\sigma$ region, which in each case points to $f_a\sim 10^9$ GeV (corresponding to a mass $m_a\sim 6\times 10^{-3}$ eV). The hinted values of $\tan\beta$ are $\gtrsim 0.7$ in DFSZ I and $\lesssim 1$ in DFSZ II.
Both have a large overlap with the region where perturbative unitarity is granted.

Finally, we see from figure \ref{fig:hinted_region_DFSZ_tanbeta_m} that the 1 $\sigma$ parameter region preferred by the
DFSZ axion interpretation of the stellar cooling anomalies can be entirely probed by IAXO+ (see appendix \ref{app:Sensitivity}), while the projected sensitivity of
ARIADNE barely touches the 1(2) $\sigma$ region for DFSZ I(II).

\subsection{Hadronic axion models}

\subsubsection{Pure KSVZ-type models}

In pure KSVZ-type axion models  \cite{Kim:1979if,Shifman:1979if}, the color anomaly arises from beyond the SM coloured vector-like fermions $Q=(Q_L,Q_R)$ transforming chirally under $U(1)_{\rm PQ}$ and according to representations $R_Q=(C_Q,I_Q,Y_Q)$ under the SM
gauge group factors $SU(3)_C\times SU(2)_L\times U(1)_Y$.  The corresponding $N$ and $E/N$ values for specific representations
$R_Q$   \cite{DiLuzio:2016sbl}
are given in the second and third column of table \ref{tab:KSVZ_fits}, respectively.
\begin{table}[h]
\begin{align}
\begin{array}{|c||c|c||c|c|c|}\hline
R_Q 					& N& E/N & f_a [{\rm GeV}]  & m_a [{\rm eV}]  & \chi^2_{\rm min}/{\rm d.o.f.}   \\	\hline \hline
(3,1,0) 				&1 & 0   & 7.7 \times 10^7 & 0.075 		  & 25.2/16 						 \\
(3,1,-\tfrac{1}{3})		&1 &2/3  & 4.9\times 10^7  & 0.12  		  & 25.2/16 						 \\
(3,2,+\tfrac{1}{6})  	&2 &5/3  & 6.4\times 10^6  & 0.88 			  & 20.9/16 				 \\
 (3,1,+\tfrac{2}{3})  	&1 &8/3  & 2.2\times 10^7  & 0.26  		  & 23.5/16 				  \\
	\hline
\end{array}
\nonumber
\end{align}
\caption{
Best fit parameters  for KSVZ-type axion interpretations of the HB, RGB, and WD cooling anomalies.
}
\label{tab:KSVZ_fits}
\end{table}

The dimensionless coupling $C_{a\gamma}$ to photons has already been shown in eq.  \eqref{eq:photon_coupling}.
At tree level, quarks and leptons do not interact with the KSVZ axion. Correspondingly, the leading contributions to the dimensionless couplings to protons and neutrons arise solely from the mixing with the pion and read~\cite{diCortona:2015ldu}:
\begin{equation}
C_{ap}^{\rm KSVZ}=-0.47(3)\,, \qquad
C_{an}^{\rm KSVZ}=-0.02(3)\,.
\end{equation}
Moreover, the dimensionless coupling to electrons is induced at one-loop level by the photon coupling and estimated as \cite{Srednicki:1985xd}
\begin{equation}
C_{ae}^{\rm KSVZ} \simeq
\frac{3\alpha^2}{2\pi}
\(
\frac{E}{N} \ln \frac{f_a}{m_e} - 1.92 \ln \frac{\Lambda}{m_e}
\)
\simeq
2.5\times 10^{-5}
\(
\frac{E}{N} \ln \frac{f_a}{m_e} - 1.92 \ln \frac{\Lambda}{m_e}
\)
\,,
\label{cagamma_ksvz}
\end{equation}
where $\Lambda \sim 1\, {\rm GeV}$.

Since $E/N$ is fixed for each KSVZ-type axion model, the only free parameter to determine the couplings
$g_{ai}^2\propto C_{ai}^2/f_a^2$ to the SM particles is $f_a$ (or, equivalently, $m_a$).
The preferred values of $f_a$ and $m_a$ from a global fit of WD+RGB+HB are given in table \ref{tab:KSVZ_fits}.

The commonly shown value of $E/N=0$ gives a rather bad fit, with $\chi^2_{\rm min}/{\rm d.o.f.}\simeq 1.6$.
A preference for $ E/N\simeq 2 $ is clearly seen. In fact, if $E/N$ were used as a fitting parameter we would find the best value to be $E/N=1.95$.
The model exploiting the representation $R_Q=(3,2,+\tfrac{1}{6})$ has $E/N=5/3$ and thus comes closest to this value, but still with a relatively bad
$ \chi^2_{\rm min}/{\rm d.o.f.}\simeq 1.3$.

The reason for this is that in KSVZ-type models the axion coupling to electrons emerges only at the loop level and is strongly suppressed with respect to tree level couplings, cf. eq. \eqref{cagamma_ksvz}.
Though the cooling anomalies indicate a preference for $C_{ae}$ about two orders of magnitude smaller than
$|C_{a\gamma}|$,  for the KSVZ axion (with $E/N$=0) the prediction for $C_{ae}$ is about an order of magnitude too small with respect to the corresponding $|C_{a\gamma}|$.
Therefore, the fit prefers a value of $E/N $ close to $1.92$ (since that lowers
$|C_{a\gamma}|$, leaving $C_{ae}$ unaffected), and increases both the couplings to photons and electrons through lower values of $f_a$, corresponding to larger values of $m_a$.
As a consequence, the models fitting better are those with  $E/N \sim 1.95$ and a large mass $\sim$ eV.

An axion in this mass range is, however, severely constrained if not totally excluded by several astrophysical and cosmological arguments.
The reader might remember that when the axion coupling to nucleons is very large, axions can interact so strongly with the medium as to remain trapped in the SN core, reducing the axion emissivity to a relatively thin and cold ``axiosphere'' that cannot compete with neutrino emission and thus does not affect the duration of the neutrino pulse of SN 1987A.
The excluded region is nominally taken to be $6\times 10^{-10}<g_{ap}<6\times 10^{-8}$, which we adapt from \cite{Raffelt:2006cw}.
On the strongly interacting side, the fewer axions emitted can interact strongly enough to be detected in Kamiokande by the nuclear reaction
$a  {}^{16}{\rm O}\to {}^{16}{\rm O}^*\to {}^{16}{\rm O}+\gamma$~\cite{Engel:1990zd}. This would exclude the range
$10^{-6}<g_{ap}<10^{-3}$  \cite{Raffelt:2006cw,Raffelt:1996wa}.
The gap $6\times 10^{-8}<g_{ap} <10^{-6}$ ($ 0.78\, {\rm eV}<m_a<13$ eV, corresponding to $0.44\times 10^6\,{\rm GeV} < f_a< 0.73\times 10^7\,$GeV) can be excluded by an independent constraint from cosmology.
In fact axions with these masses are produced thermally in the early universe and constitute a sizeable hot dark matter component, in analogy to massive neutrinos.
Cosmological precision data provide an upper limit on the possible hot dark matter
fraction which translates into an upper limit on the axion mass, $m_a<0.8$ eV at 95\%C.L.~\cite{Hannestad:2010yi,Archidiacono:2013cha,DiValentino:2015wba}, which in principle closes the gap.\footnote{In the DFSZ cases, a small $f_a$ solution in the SN 1987A-Kamiokande gap is excluded by the $R$-parameter, which is sensitive to the $g_{a\gamma}$ coupling.}
The search for monochromatic photon lines from axion dark matter decay excludes even more strongly the region $4.5\,{\rm eV} <m_a <7.7\,{\rm eV}$ \cite{Grin:2006aw}, see also \cite{Cadamuro:2011fd}.
Once more we warn the reader that SN 1987A constraints have uncontrolled systematic uncertainties themselves.
The numbers presented here have to be taken with a grain of salt and as an invitation to review the situation once more precise modeling is available.
Note finally that we have not considered the hint from the NS in CAS A, since
the interaction of the KSVZ axion with the neutron is compatible with zero.

We conclude that a KSVZ-type axion can not give a plausible explanation of the cooling anomalies without severely compromising other astrophysical constraints.

\subsubsection{KSVZ-type axion/majoron}
\label{sec:KSVZ-type axion/majoron}

In refs.  \cite{Shin:1987xc,Dias:2014osa,Ballesteros:2016euj,Ballesteros:2016xej},
the KSVZ-like particle content is extended by three right-handed SM-singlet (``sterile") neutrinos $N_{Ri}$,  and the Peccei-Quinn (PQ) symmetry is unified with the lepton symmetry, explaining the smallness of the masses of the active neutrinos by the seesaw mechanism~\cite{Minkowski:1977sc,GellMann:1980vs,Yanagida:1979as,Mohapatra:1979ia}. The model then features  Yukawa couplings to the left-handed SM lepton doublets $L$, the SM Higgs
doublet $H$, and the PQ complex scalar $\sigma$ as follows:
\begin{align}
{\mathcal L}_Y \supset  -\overline{L}_i F_{ij} N_{Rj} H -\tfrac12\overline{N}_{Ri}^c Y_{ij}  N_{Rj} \sigma + {\rm h.c.}\, ,
\label{eq:lagrangian}
\end{align}
where $F_{ij}$ and $Y_{ij}$ are Yukawa matrices.
PQ symmetry breaking leads to the Majorana mass matrix $M_M =  Y v_{\rm PQ}/\sqrt2$ which can be taken diagonal without loss of generality.
Electroweak symmetry breaking, $\langle H\rangle = (v/\sqrt2,0)^T$, introduces a mixing between the left and right-handed neutrinos via the Dirac mass matrix $m_D = F v/\sqrt2$. The full Majorana mass matrix, in the basis $(\nu^c, N_R) = V n$, is then
\begin{align}
\matrixx{0 & m_D \\ m_D^T & M_M} = V^* \diag (m_1,\dots, m_6) V^\dagger\,,
\label{eq:neutrino_mass_matrix}
\end{align}
where $V$ is the $6\times 6$ mixing matrix to the states $n_R$, which form the Majorana mass eigenstates $n = n_R + n_R^c$.

In these models,
the axion $A$ is at the same time the majoron  $J$ -- the pseudo Nambu-Goldstone boson
arising from spontaneous breaking of the lepton symmetry \cite{Chikashige:1980ui}. Importantly, the one-loop induced axion-electron coupling,
\begin{equation}\label{Eq:axion_majoron_Cae}
C_{ae}^{{\rm KSVZ}\ A/J} \simeq  C_{ae}^{\rm KSVZ} + C_{ae}^{A/J}
\end{equation}
gets an extra contribution from the loop involving the sterile neutrinos $N_{Ri}$ \cite{Shin:1987xc,Pilaftsis:1993af}, which -- to lowest order in the seesaw limit,  $m_D/M_M\ll 1$ -- is given by \cite{Garcia-Cely:2017oco}
\begin{align}
\label{CaeAXIMAJ}
C_{ae}^{A/J} &\simeq -\frac{1}{16\pi^2 N} \left(  \tr \kappa -2 \kappa_{ee}   \right)
 \,,
\end{align}
where the dimensionless hermitian $3\times 3$ matrix $\kappa$ is defined as
\begin{align}
\kappa \equiv  \frac{m_D m_D^\dagger}{v^2} = \frac{F F^\dagger}{2}\, .
\label{eq:K}
\end{align}
We note that all the diagonal entries of $\kappa$ are real and non-negative, and thus
its trace is non-negative \cite{Garcia-Cely:2017oco},
	\begin{align}
\tr \kappa \geq 3\, (\det \kappa )^{1/3} =
	3\, \left( \frac{1}{v^6}\prod_{j=1}^6 m_j \right)^{1/3} \geq  0 \, .
	\end{align}
In addition, the diagonal entries are bounded from above by perturbative unitarity, so
\begin{equation}
0 \leq \kappa_{\ell \ell} \lesssim 4\pi \,.
\end{equation}

\begin{figure}[t]
\centering
\hspace{-.5cm}
\includegraphics[width=0.4\textwidth]{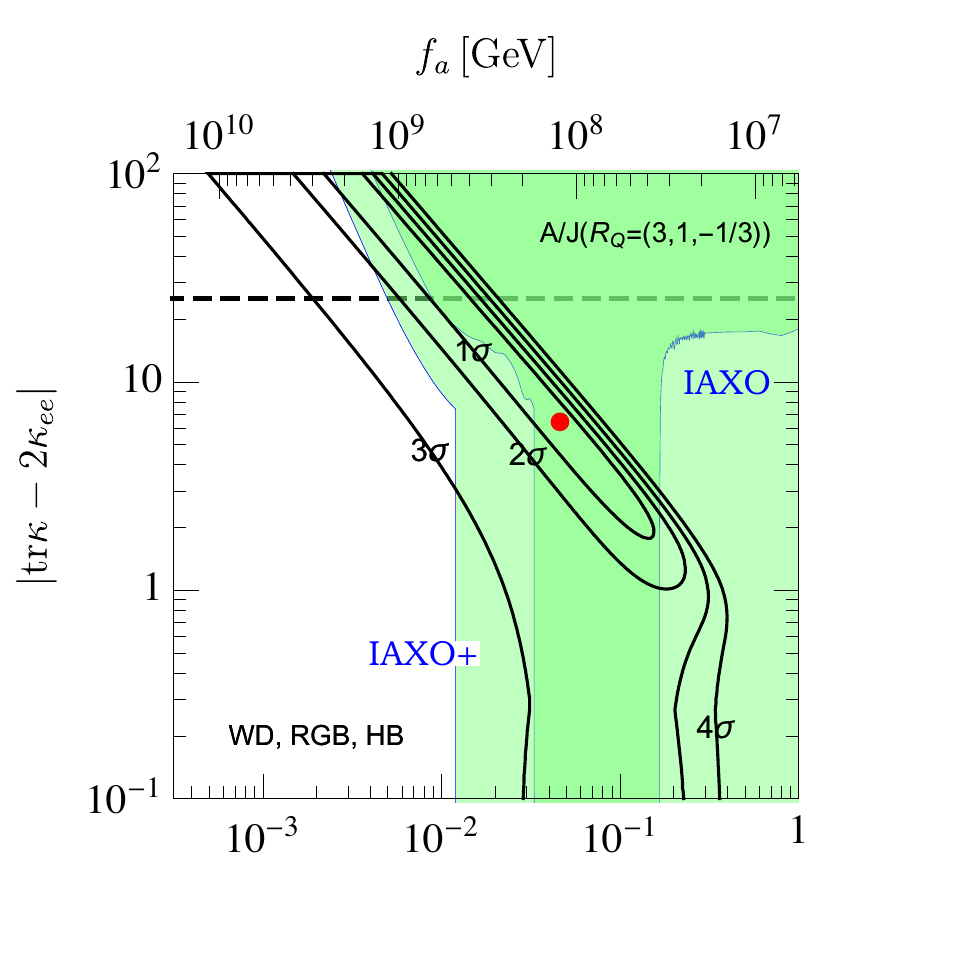}
\hspace{-2cm}
\includegraphics[width=0.4\textwidth]{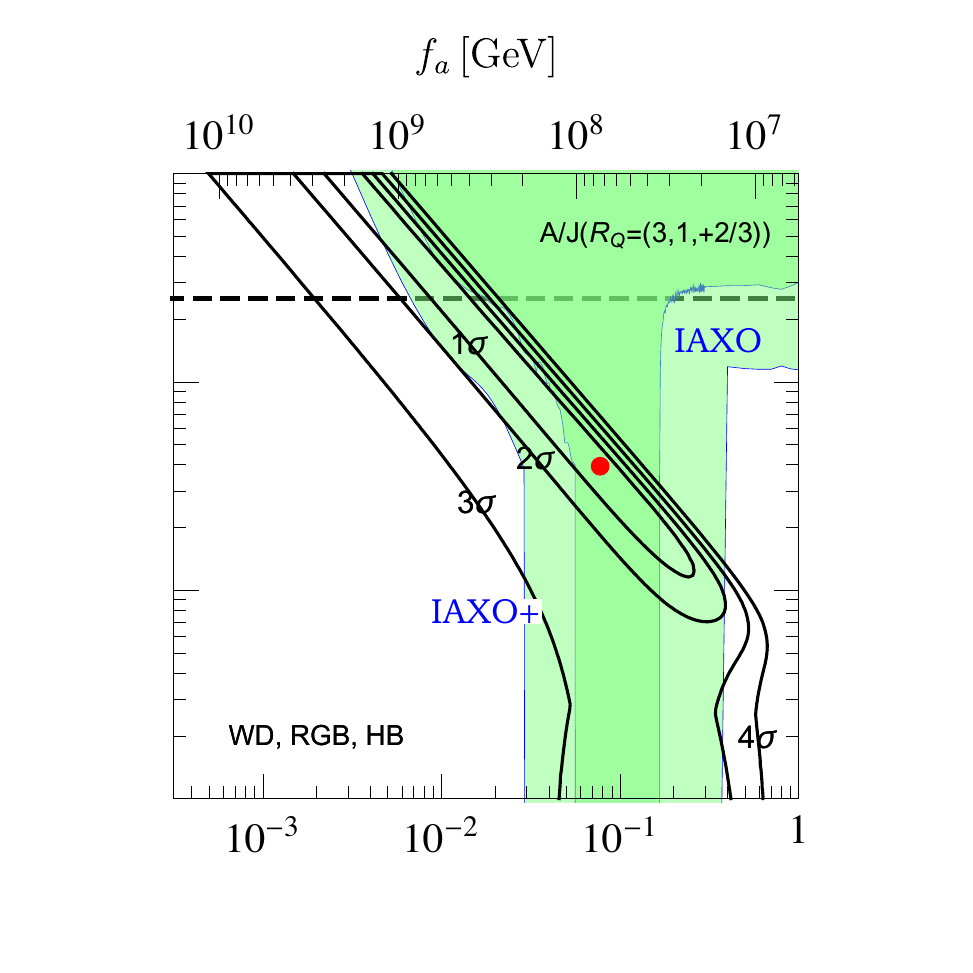}
\hspace{-2cm}
\includegraphics[width=0.4\textwidth]{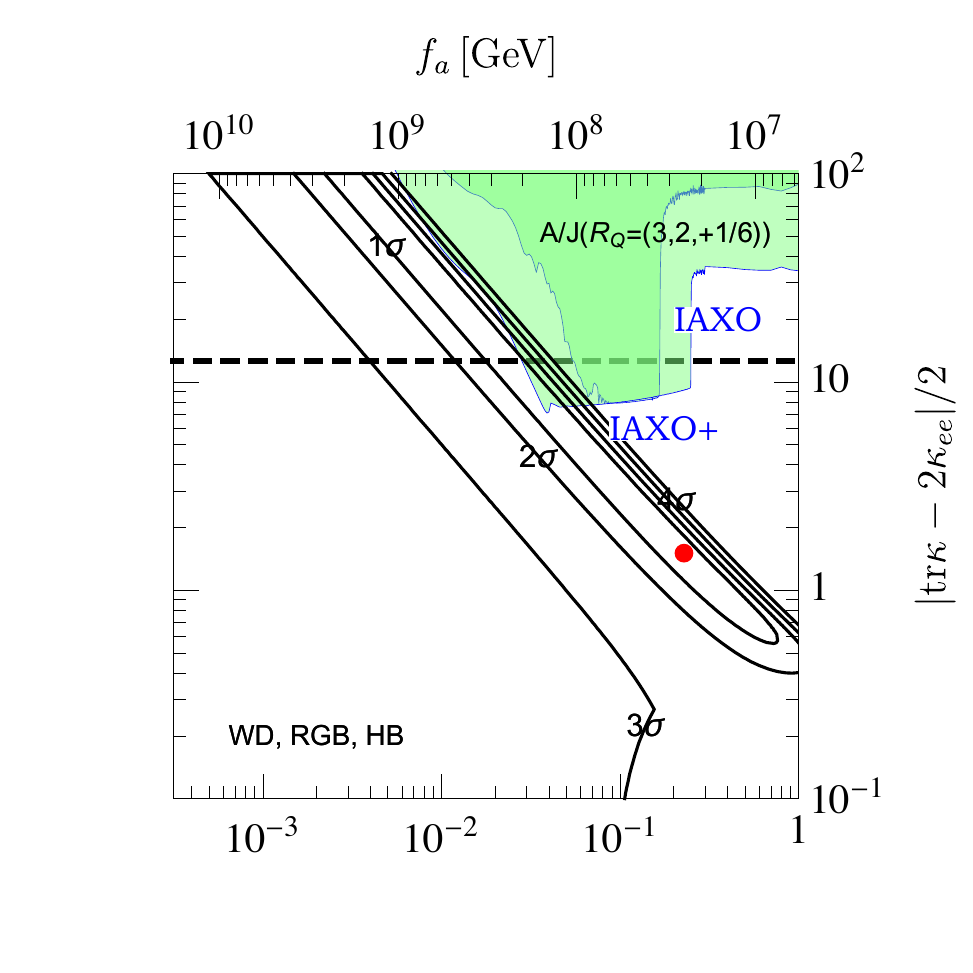}
\vspace{-1cm}

\hspace{-.5cm}
\includegraphics[width=0.4\textwidth]{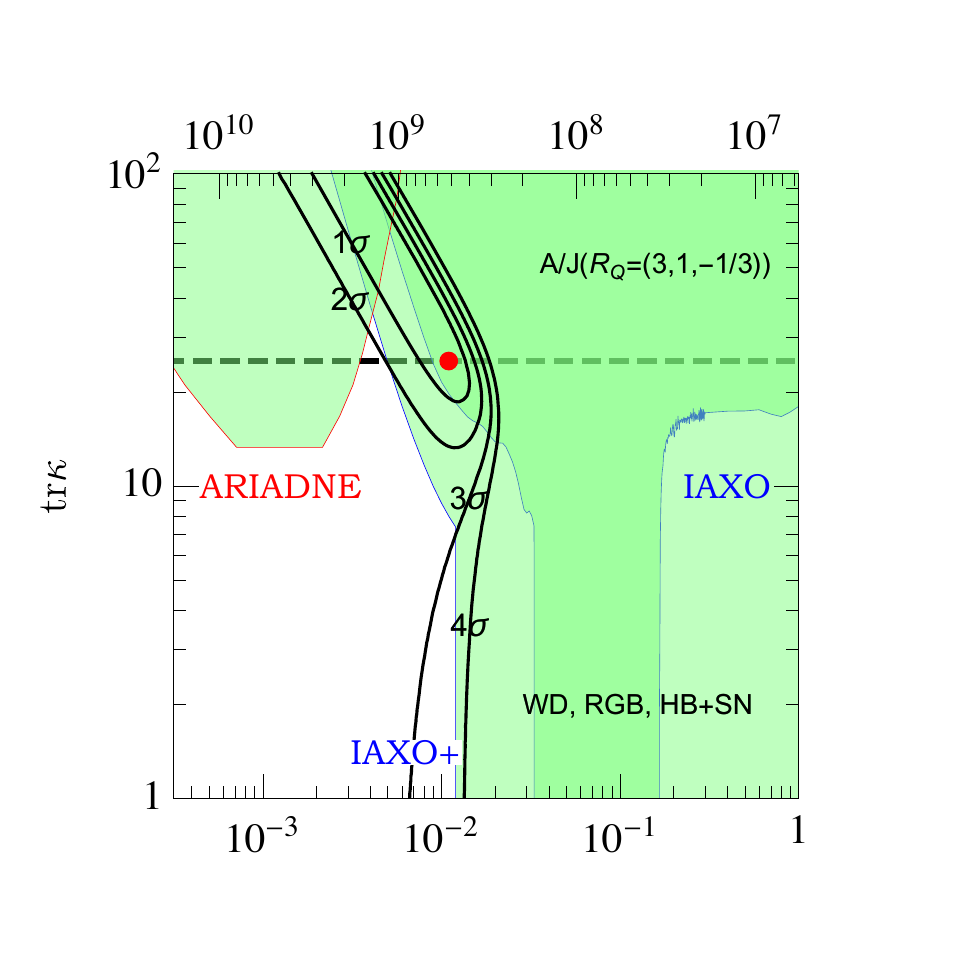}
\hspace{-2cm}
\includegraphics[width=0.4\textwidth]{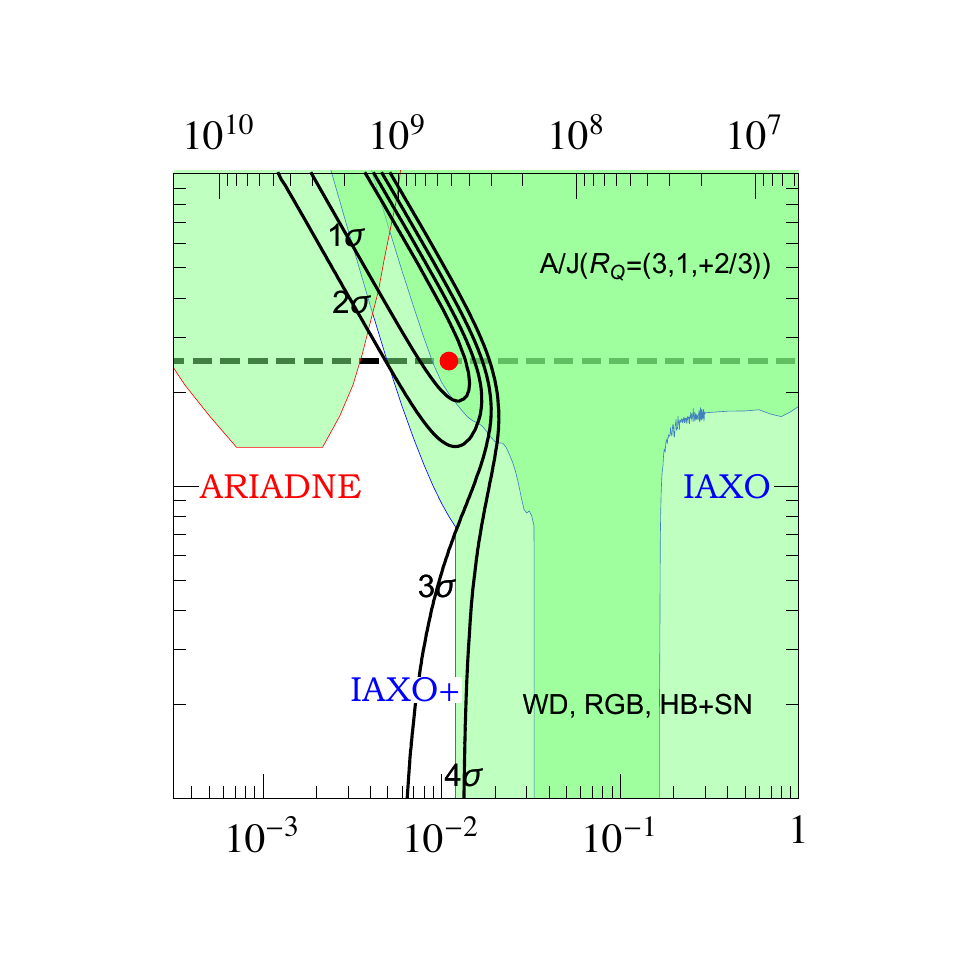}
\hspace{-2cm}
\includegraphics[width=0.4\textwidth]{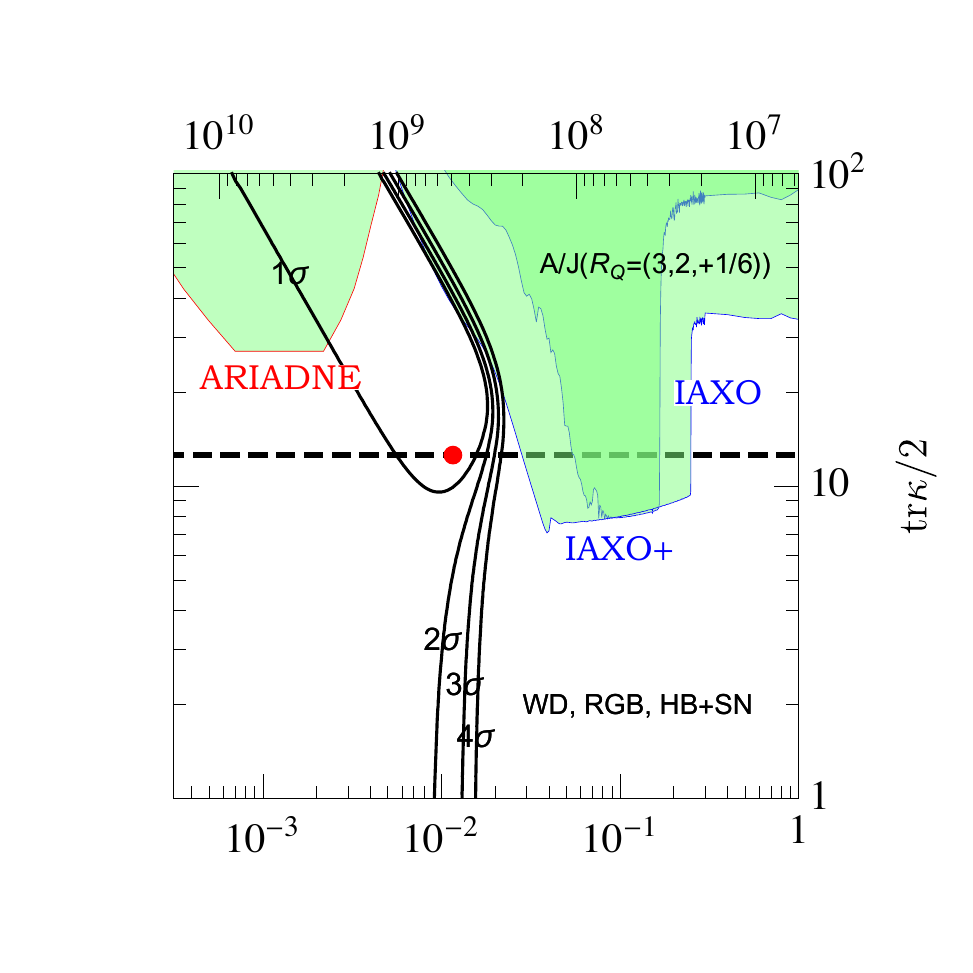}
\vspace{-1cm}

\hspace{-.5cm}
\includegraphics[width=0.4\textwidth]{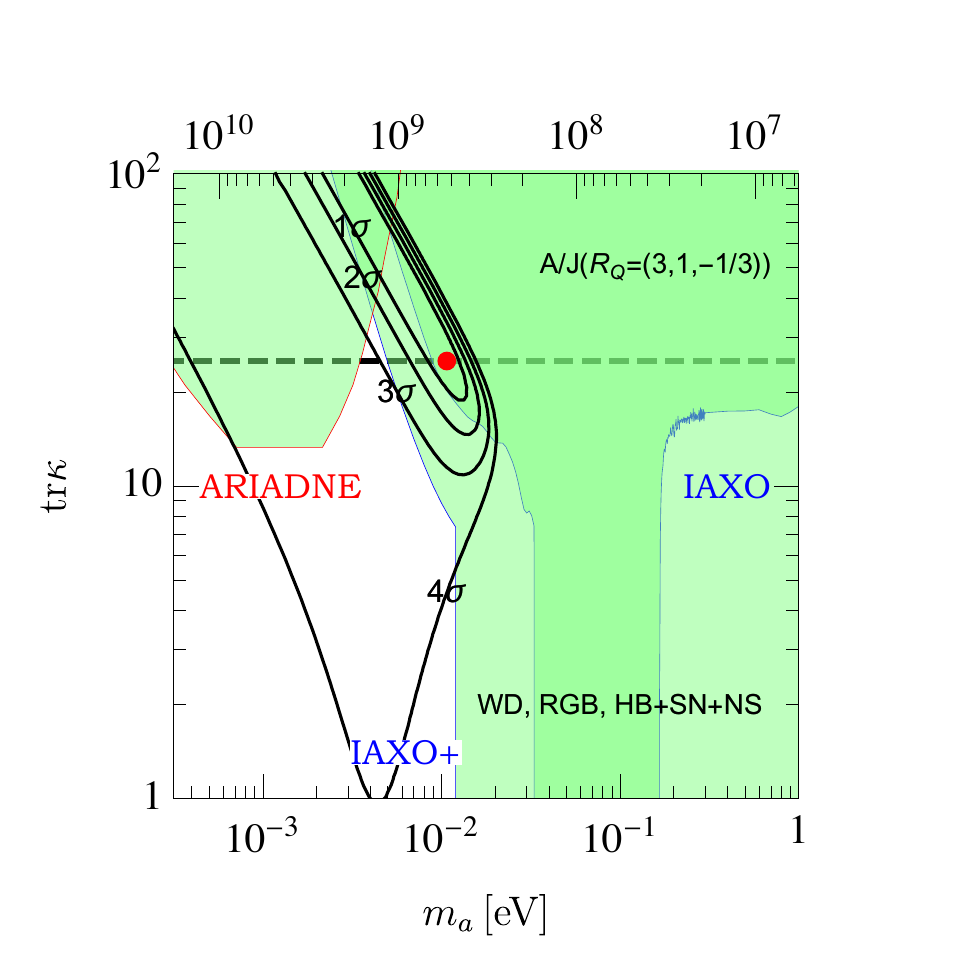}
\hspace{-2cm}
\includegraphics[width=0.4\textwidth]{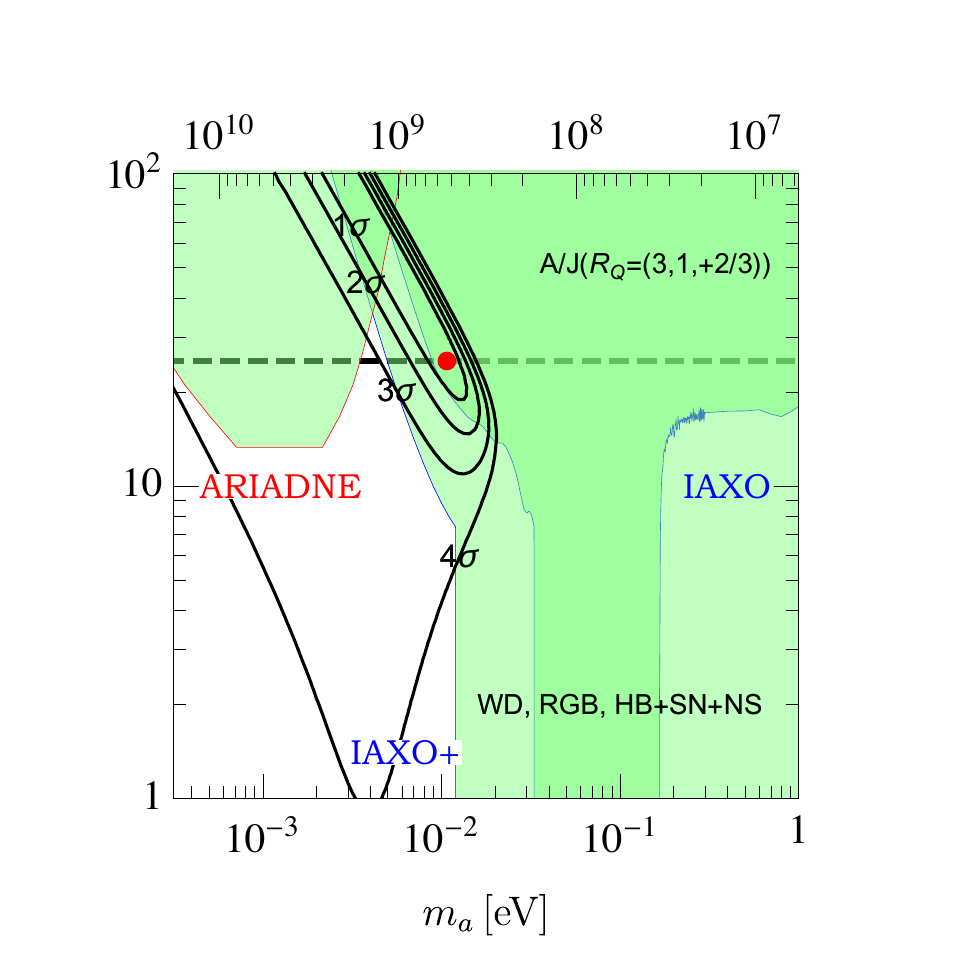}
\hspace{-2cm}
\includegraphics[width=0.4\textwidth]{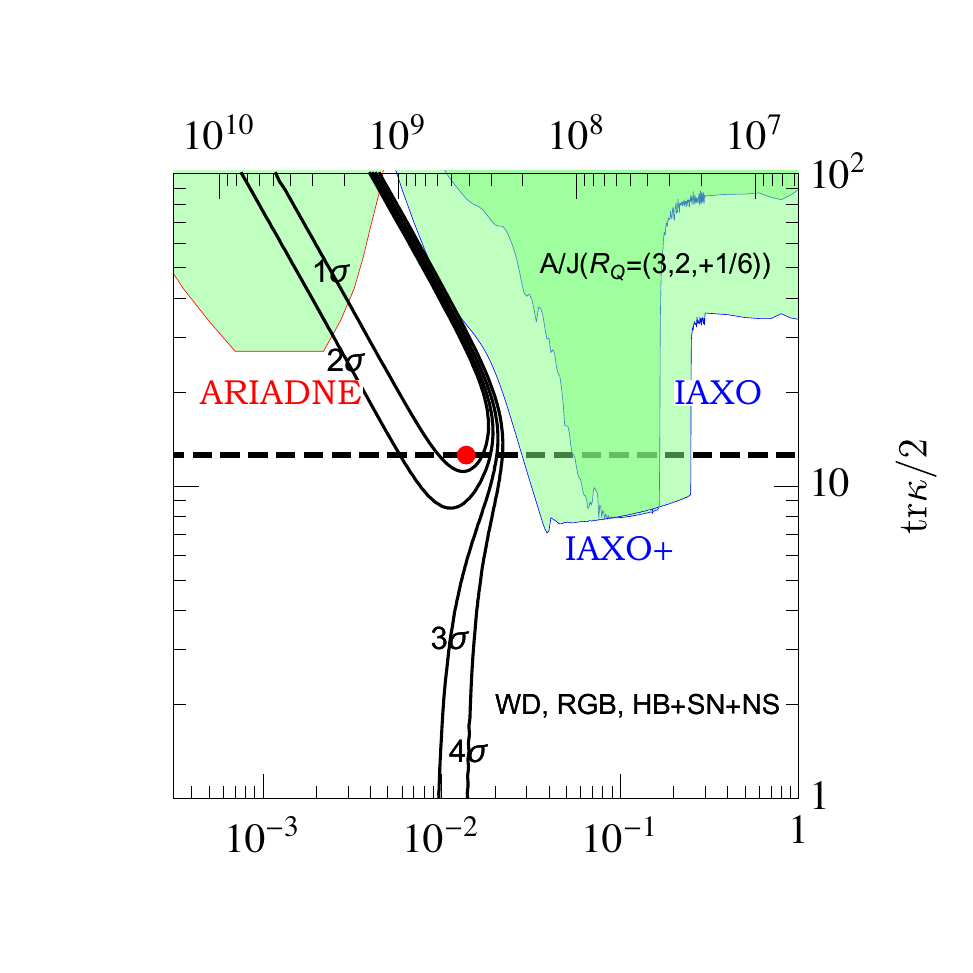}
\vspace{-.5cm}

\caption{
1, 2, 3, 4\,$\sigma$ contours in the global fit of KSVZ $A/J$ models (left panels for $R_Q=(3,1,-\tfrac{1}{3})$;
middle panels for $R_Q=(3,1,+\tfrac{2}{3})$; right panels for $R_Q=(3,2,+\tfrac{1}{6})$) with WD+RGB+HB data alone (top) and including the SN 1987A constraint (middle) and, in addition,  the NS CAS A data (bottom).
Perturbative unitarity for the Dirac Yukawas $F$ is satisfied for values of  $\tr \kappa$  below the dashed lines.
The red bullets are the best fits compatible with perturbative unitarity.
Also shown are the projected sensitivities of ARIADNE \cite{Arvanitaki:2014dfa}  and IAXO~\cite{Armengaud:2014gea} (see appendix \ref{app:Sensitivity}).}
\label{fig:hinted_region_KSVZ_AJ_all}
\end{figure}

In KSVZ $A/J$, the dimensionless axion-electron coupling can be larger than
in the KSVZ model, as evident from eq. \eqref{CaeAXIMAJ}.
In particular, for $\left| \tr \kappa -2\kappa_{ee}  \right|/N={\mathcal O}(1)$, it is naturally of order
$C_{ae}^{A/J}={\mathcal O}(10^{-2})$, as favoured by the best fit of the stellar cooling hints
from WDs, the RGB, and HBs (for $f_a = {\mathcal O}(10^{8})$\,GeV).
Best fit values to reproduce the WD+RGB+HB cooling anomalies with $\left| \tr \kappa -2\kappa_{ee}  \right|/N$ and $f_a$ are shown in table \ref{tab:SMASH_fits_wd_rg_hb} for 3  simple $A/J$ models.
The fits are very good, with $\chi^2_{\rm min}/{\rm d.o.f.} =1.0$.
The 1, 2 and 3 $\sigma$ contours in the $\left| \tr \kappa -2\kappa_{ee}  \right|$ vs $f_a$ plane are shown in  figure \ref{fig:hinted_region_KSVZ_AJ_all} (top panel).
The fit is once more dominated by the WD+RGB that force $g_{ae}\sim 1.5 \times 10^{-13}$,  imposing a degeneracy between
$\left| \tr \kappa -2\kappa_{ee}  \right|$ and $f_a$. This results in the diagonal shape of the contours. The degeneracy is broken at small $f_a$ when
$g_{a\gamma}$ becomes large and starts to ruin the $R$-parameter fit.
In the  $R_Q=(3,2,+\frac{1}{6})$ case, the $C_{a\gamma}$ is smaller, so the 1 $\sigma$ best fit region extends further in $m_a$,
almost reaching $m_a \sim $1 eV.
The best fits prefer sizeable diagonal matrix elements in $\kappa$, of the order of the perturbative unitarity bound $\left| \tr \kappa -2\kappa_{ee}  \right|\gtrsim 1$.
At the smallest values of $\left| \tr \kappa -2\kappa_{ee}  \right|\sim 0.1 E$ the photon-loop contribution to $C_{ae}$ produces a visible feature because it can cancel completely the RH neutrino contribution.

\begin{table}[h]
\begin{align}
\begin{array}{|c||c|c||c|c|c|c|}\hline
R_Q 					& N& E/N & f_a\, [10^9\,{\rm GeV}]  & m_a [{\rm meV}]  & | \tr \kappa -2\kappa_{ee}|  &  \chi^2_{\rm min}/{\rm d.o.f.}
\\	\hline \hline
(3,1,-\tfrac{1}{3})	&1 &2/3  & 0.12   & 46 	& 6.2  & 14.9/15
\\
 (3,1,+\tfrac{2}{3})  	&1 &8/3  & 0.073 & 77 	& 3.7    & 14.9/15
\\
(3,2,+\tfrac{1}{6})  	&2 &5/3  & 0.025 & 230 	& 2.5  & 14.9/15
\\
	\hline
\end{array}
\nonumber
\end{align}
\caption{
Best fit parameters compatible with  perturbative unitarity for KSVZ $A/J$ interpretations of the HB, RGB, and WD cooling anomalies.
}
\label{tab:SMASH_fits_wd_rg_hb}
\end{table}

The KSVZ $A/J$ may also have a larger coupling to the neutron than the pure KSVZ axion,
due to the loop induced contribution to the axion-quark couplings from the sterile neutrinos,
\begin{align}
C_{aq}^{A/J} &\simeq \frac{1}{8\pi^2 N}  \, T^{q}_3\, \tr \kappa  \, ,
\end{align}
with $T^{d}_3 = -\tfrac12 = - T^u_3$.
This contribution leads to a non-zero axion coupling to the neutron, $C_{an}$, and thus may explain the NS cooling hint.
For completeness we include the corrections to the axion proton coupling to consider the SN 1987A constraint as well. We find,
\begin{align}
\label{eq:nucleon_couplings_ksvz}
C^{{\rm KSVZ}\ A/J}_{ap} & = -0.47(3)+
\frac{1}{16\pi^2 N}\left[ 0.88(3) -0.012(5) -0.0035(4) \right] \tr \kappa   \nonumber \\
				&\hspace{2.15cm}  +\frac{1}{16\pi^2 N}\left[ 0.39(2) + 0.038(5) + 0.009(2) \right] \tr \kappa \,, \nonumber  \\
C^{{\rm KSVZ}\ A/J}_{an} & = -0.02(3)-\frac{1}{16\pi^2 N}\left[ 0.88(3)-0.038(5) -0.009(2) \right] \tr \kappa  \nonumber \\
				&\hspace{2.15cm}-\frac{1}{16\pi^2 N}\left[0.39(2) + 0.012(5)+0.0035(4)\right] \tr \kappa \,.
\end{align}

The inclusion of the SN 1987A constraint in the global fit has dramatic consequences (see middle panels in figure \ref{fig:hinted_region_KSVZ_AJ_all}).
In this case, in general,  three $A/J$ parameters have to be fitted: $f_a$, $\left|\tr \kappa-2 \kappa_{ee}    \right|$, and $\tr \kappa$.
It turns out that our minima of $\chi^2$ happen always at $\kappa_{ee}=0$, so that
using $\kappa_{ee}=0$ is equivalent to marginalising over it.
Moreover, for hierarchical values of the diagonal elements of $\kappa$, axion couplings effectively depend only on $\tr \kappa$. In fact, for
$\kappa_{\mu\mu}$, $\kappa_{\tau\tau}\gg \kappa_{ee}$ and $\kappa_{ee}\gg \kappa_{\mu\mu}$, $\kappa_{\tau\tau}$, we have
$|\tr \kappa-2 \kappa_{ee}| \approx  \tr \kappa$.
Note that the unitarity limit of tr$\kappa$ is $8\pi$ and $4\pi$ respectively in these limits.
For simplicity we will assume $|\tr \kappa-2 \kappa_{ee}|\sim |\tr \kappa|$ in the following and display the most conservative $8\pi$ unitarity limit.

Axion masses above $m_a\sim$ 10 meV become excluded by the SN argument and so
the 1 $\sigma$ region recedes towards the region inaccessible to reliable perturbative calculations.
All the best fits have smaller $\chi^2_{\rm min}$/d.o.f. We show the values for the best case $R_Q=(3,1,-\frac{1}{3})$ in
table \ref{tab:SMASH_fits_wd_rg_hb_plus}.
The worst case corresponds to $R_Q=(3,2,+\frac{1}{6})$, because for $N=2$ the maximum electron coupling compatible with perturbative unitarity is half the value than in the $N=1$ cases.
We see however that the favoured regions correspond to the highest masses $m_a\sim 10-20$ meV and that there is a sizeable region of parameters within the 1 $\sigma$ region, although one has to remember that the best fit is not so good.

\begin{table}[h]
\begin{align}
\begin{array}{|l||c|c|c|c|}\hline
{\rm Global\ fit\ includes}					&  f_a\, [10^9\, {\rm GeV}]  & m_a [{\rm meV}]  & \tr \kappa   &  \chi^2_{\rm min}/{\rm d.o.f.}
\\	\hline \hline
\rm WD,RGB,HB,SN  		& 0.52  & 11 	&	8\pi	 & 16.9/16 						
\\
\rm WD,RGB,HB,NS,SN	& 0.53  & 11  	&	8\pi  	& 17.0/17	
\\
	\hline
\end{array}
\nonumber
\end{align}
\caption{
Best fit parameters compatible with  perturbative unitarity for  KSVZ $A/J$   interpretations of the  cooling anomalies ($R_Q=(3,1,-\tfrac{1}{3})$).
}
\label{tab:SMASH_fits_wd_rg_hb_plus}
\end{table}

Adding the hint from the NS in CAS A actually improves the fit,  see table \ref{tab:SMASH_fits_wd_rg_hb_plus}, so that the preferred regions shrink a bit, cf. figure \ref{fig:hinted_region_KSVZ_AJ_all} (bottom panel).
This is due to a remarkable prediction of $A/J $ models. Neglecting the $-0.02(3)$ factor in $C_{an}$ we have
\begin{equation}
\left|\frac{C_{an}}{C_{ae}}\right|=1.24 \frac{{\rm tr \kappa}}{|{\rm tr \kappa}-2\kappa_{ee}|}
\end{equation}
while the ratio of the neutron and electron couplings favoured by the CAS A NS and the combination WD+RGB+HB, respectively, is
\begin{equation}
\left|\frac{C_{an}}{C_{ae}}\right|=\frac{g_{an} /m_n}{g_{ae}/m_e}=\frac{3.8\times 10^{-10}}{1.5\times 10^{-13}}\frac{m_e}{m_n}=1.4
\end{equation}
Therefore, $A/J$ models that fit the combination WD+RGB+HB tend to give a good fit to NS
unless $2\kappa_{ee}\sim \kappa_{\mu\mu}+\kappa_{\tau\tau}$, the only case where the ratio
$\frac{{\rm tr \kappa}}{|{\rm tr \kappa}-2\kappa_{ee}|}$ can differ significantly from 1.

For $R_Q=(3,1,-\tfrac{1}{3})$ and $R_Q=(3,2,+\tfrac{1}{6})$, the 1 $\sigma$ parameter region preferred by the KSVZ $A/J$  interpretation of the stellar cooling anomalies can be entirely probed by IAXO+, while
the case $R_Q=(3,2,+\tfrac{1}{6})$ lies outside of sensitivity of IAXO. ARIADNE has sensitivity only in the region outside of perturbative control,
cf. figure \ref{fig:hinted_region_KSVZ_AJ_all}.

\section{Axion hints compatible with axion dark matter?}
\label{sec:dm}

In addition to providing a possible interpretation of the stellar cooling hints, the axion can also be a good candidate of dark matter.
Although the mass range obtained by global fits to the stellar cooling hints is much higher than the conventional one~\cite{Preskill:1982cy,Abbott:1982af,Dine:1982ah},
there is a possibility that the axion becomes the main constituent of dark matter at those large masses.
This possibility arises from the fact that the relic axion abundance strongly depends on the early history of the universe.

We consider two possibilities in the context of PQ symmetry breaking in inflationary cosmology.
One possibility is to assume that the $U(1)_{\rm PQ}$ symmetry has been broken before inflation and is never restored.
In this scenario, the present observable universe is predicted to be in the same vacuum due to inflation homogenising the universe up to scales
beyond the horizon today. We can easily assume that the few domain walls present in the universe are not in our local observable patch.
In this case, axions are produced by the standard vacuum re-alignment mechanism~\cite{Preskill:1982cy,Abbott:1982af,Dine:1982ah}
with the initial axion angle $\theta_i = \langle a\rangle /f_a$ fixed at the inflationary epoch.
The prediction of the relic axion abundance depends not only on $f_a$, but also on $\theta_i$.
In particular, the slow-roll behavior of the axion field in the limit $\theta_i\to \pi$
opens up a possibility that the axion becomes the main constituent of dark matter in the $m_a<$ meV range~\cite{Wantz:2009it}.
A potential problem with this scenario is that the axion dark matter imprints isocurvature fluctuations in the cosmic microwave background temperature anisotropies
and these are enhanced in the $\theta_i\to \pi$ limit.
The null observation of the isocurvature fluctuations by the Planck mission leads to a severe upper bound on the Hubble expansion rate during inflation,
$H_{\rm inf} \lesssim\mathcal{O}(100)\,\mathrm{GeV}$ for $f_a = 10^{10}\,\mathrm{GeV}$,
and it becomes even tighter for $\theta_i\to \pi$~\cite{Kobayashi:2013nva}.
Therefore, it is unlikely that the axion whose mass fits to the stellar cooling hints
becomes the main constituent of dark matter in the non-restored symmetry scenario.

Next, we consider the alternative scenario in which the $U(1)_{\rm PQ}$ symmetry is restored after inflation.
In this case, some class of axion models suffer from a serious cosmological domain wall problem~\cite{Sikivie:1982qv}.
In general, the low energy effective potential for the axion field has $N$ degenerate minima connected by $N$ maxima that lead to
$N$ different type of domain wall solutions. Hence, $N$ is sometimes called domain wall number and denoted as $N_{\rm DW}$.
In the DFSZ models we have $N_{\rm DW}=6$, while in KSVZ-type axion models, the value of $N_{\rm DW}$ depends on the representation
$R_Q$ and the number $N_Q$ of exotic quarks introduced in the theory.
For instance, $N_{\rm DW}= N_Q$ if we take $R_Q = (3,1,-\frac{1}{3})$ or $R_Q = (3,1,+\frac{2}{3})$
and assume that PQ charges of $Q$'s have the same sign,
while $N_{\rm DW} = 2N_Q$ if we take $R_Q = (3,2,+\frac{1}{6})$ and assume that PQ charges of $Q$'s have the same sign.
If the different vacua are populated in the early universe, domain walls are created as boundaries between them.
For $N_{\rm DW}>1$, the network of domain walls is stable and its energy density redshifts slower than that of radiation~\cite{Vilenkin:2000jqa}, and hence it tends to dominate the total energy density of the universe,
causing trouble with the standard cosmology~\cite{Zeldovich:1974uw}.
There is no domain wall problem if $N_{\rm DW}=1$, because the network is unstable in this case \cite{Vilenkin:1982ks}.

A domain wall number equal to one is realized for example
in KSVZ-type models with $N_Q=1$ representation of $R_Q = (3,1,-\frac{1}{3})$ or $R_Q = (3,1,+\frac{2}{3})$. However, in this case, the required
axion mass to explain the observed dark matter abundance is predicted to be in the range
$50\,\mu{\rm eV}\lesssim m_a\lesssim 200\,\mu{\rm eV}$ \cite{Hiramatsu:2012gg,Kawasaki:2014sqa,Ballesteros:2016xej} ---
way too small to explain the stellar cooling anomalies\footnote{It is argued that the theoretical uncertainty of the mass of axion dark matter in the scenario with $N_{\rm DW}=1$ might be even larger because of an effect overlooked in the previous literature~\cite{Fleury:2015aca}. A recent alternative analysis tentatively finds even smaller mass values, by a factor of order one~\cite{Fleury:2016xrz}.}.

Therefore, we have to consider the alternative scenario with
domain wall number larger than one, which is automatic in DFSZ-type models and can be arranged in the simplest KSVZ-type models by taking $N_Q>1$.
In this case, the domain wall problem can be avoided if
new physics breaks the degeneracy between the different vacua~\cite{Gelmini:1988sf,Larsson:1996sp}.
Such energy difference between vacua creates pressure that
renders the domain walls unstable.
Here we assume that the $U(1)_{\rm PQ}$ symmetry is explicitly broken due to the following Planck-suppressed operator~\cite{Ringwald:2015dsf}
\begin{equation}
\mathcal{L} \supset
g M_{\rm Pl}^{4} \(\frac{\sigma}{M_{\rm Pl}}\)^{\cal N} + \mathrm{h.c.},
\end{equation}
where $\sigma$ is the SM singlet complex scalar field, $M_{\rm Pl}\simeq 2.435\times 10^{18}\,\mathrm{GeV}$ is the reduced Planck mass,
${\cal N}$ is an integer\footnote{In ref.~\cite{Ringwald:2015dsf},  the PQ symmetry is assumed to arise as an accidental symmetry from a discrete $Z_{\cal N}$ symmetry.}, and $g$ is a complex dimensionless constant.
In the low energy effective Lagrangian, the above operator induces an additional potential for the axion field
\begin{equation}
V_{g} =
-2|g| M_{\rm Pl}^{4} \(\frac{v_{\rm PQ}}{\sqrt{2}M_{\rm Pl}}\)^{\cal N} \cos\left({\cal N}\frac{a}{v_{\rm PQ}}+\Delta\right)
\label{axion_potential_Vg}
\end{equation}
which lifts $N_{\rm DW}$ degenerate vacua and induces late-time annihilation of domain walls.
Here, $\Delta$ includes the phase of coupling $g$,
\begin{equation}
\Delta=\mathrm{arg}(g)- \frac{{\cal N}}{N_{\rm DW}}\bar{\theta},
\end{equation}
and $\bar{\theta}$ is the sum of the QCD $\theta$ parameter and the contribution from the quark mass phases
(we have redefined the axion field to have $\langle a\rangle =0$ at the CP conserving minima).

The collapse of long-lived domain walls produces an additional number of axions,
that contribute to and can be the dominant component of the dark matter density~\cite{Hiramatsu:2012sc,Kawasaki:2014sqa}.
In this case, the relic axion abundance depends not only on $f_a$, but also on
the explicit symmetry breaking parameter $g$ and on the order of the Planck-suppressed operator ${\cal N}$.
Here we follow the analysis in refs.~\cite{Kawasaki:2014sqa,Ringwald:2015dsf} and find that the hinted mass ranges obtained in the previous section
can be compatible with axion dark matter if ${\cal N}=9$.
We note that, for ${\cal N}\le 8$, the axion mass is determined by $V_g$ in eq.~\eqref{axion_potential_Vg}
rather than the potential induced by the topological susceptibility in QCD unless $|g|$ takes an extremely small value~\cite{Ringwald:2015dsf}.
Furthermore, we find that the observed dark matter abundance cannot be explained with $|g| < 1$ if ${\cal N}\ge 10$.

In table~\ref{tab:axion_DM_parameters}, we summarise the required values of $|g|$ that allow fitting the observed cold dark matter abundance
for benchmark values of the axion mass obtained in the previous section.
The benchmarks are the best fit values compatible with  perturbative unitarity.
For KSVZ $A/J$ models, we consider three further choices of the domain wall number, $N_{\rm DW} = 2$, $4$, and $6$ with the representation $R_Q = (3,1,-\frac{1}{3})$ by simply adding $N_{\rm DW}$ generations in the same representation.
These models are always at the border of the perturbative unitarity region, and have increasing $\chi^2_{\rm min}$/d.o.f. as we increase $N_{\rm DW}$.
The tension with unitarity and the SN constraint becomes so severe at large $N_{\rm DW}$ that
the best fits move towards $g_{ae}<1.5\times 10^{-13}$ worsening the fit to WD+RGB+HB, because the likelihood decreases slower in that direction than in the small $f_a$ direction, where it faces the SN constraint.

Note that for all the models the axion decay constant takes a larger value when we include the SN bound, which increases the energy bias in our parametrisation.
Therefore, a smaller value of $|g|$ is required in order to compensate it.

\begin{table}[h]
\begin{align}
{\footnotesize
\begin{array}{|l|c||l|c|c|c|c|}\hline
\text{Model} & N_{\rm DW}  & \text{Global fit includes} & f_a [10^8\, {\rm GeV}]  & m_a [{\rm meV}] &  |g| & \text{Upper limit on}\ \Delta \\
\hline \hline
\text{DFSZ I} & 6 & \text{HB,RGB,WD} 	& 0.77 & 74 	& {0.033 \textendash 1.8} & {1.1 \textendash 1.5} \\
 & 6 & \text{HB,RGB,WD,NS,SN}          	& 9.9 & 5.8 	& {(0.72 \textendash 6.2)\times 10^{-8}} & {(0.26 \textendash 2.4)\times 10^{-2}} \\
 \hline
\text{DFSZ II} & 6 & \text{HB,RGB,WD} 	& 1.2 & 47  	& {(0.23 \textendash 5.7)\times 10^{-2}} & {0.52 \textendash 1.5} \\
 & 6 & \text{HB,RGB,WD,NS,SN}          	& 9.1 & 6.3  	& {(0.12 \textendash 1.1)\times 10^{-7}} & {(0.32 \textendash 2.9)\times 10^{-2}} \\
\hline
		& 2 &\text{HB,RGB,WD}& 1.2 	& 46  & {(0.031 \textendash 2.8)\times 10^{2}} & {0.74 \textendash 1.6} \\
	& 4 & 			     & 	1.2	& 46  & {0.064 \textendash 4.1} & {0.21 \textendash 2.3} \\
 \text{KSVZ $A/J$} 				  	& 6 &			     & 0.86  & 66 & {(0.17 \textendash 7.1)\times 10^{-1}} & {0.93 \textendash 1.5} \\
	R_Q=(3,1,-\frac{1}{3}) 	 			& 2 & \text{HB,RGB,WD,NS,SN}
									     & 4.1 & 14 & {(0.19 \textendash 7.6)\times 10^{-2}} & {(0.39 \textendash 9.9)\times 10^{-1}} \\
 					& 4 & 			     & 6.1 & 9.2 & {(0.049 \textendash 1.0)\times 10^{-4}} & {(0.31 \textendash 6.6)\times 10^{-2}} \\
					& 6 & 			     & 8.0 & 7.1 & {(0.26 \textendash 2.5)\times 10^{-7}} & {(0.44 \textendash 4.2)\times 10^{-2}} \\
\hline
\end{array}
}
\nonumber
\end{align}
\caption{
Values of the symmetry breaking parameters $|g|$ and $\Delta$ (for the case ${\cal N}=9$) that explain the present dark matter abundance
for different benchmark axion models explaining the stellar cooling anomalies.
The range of values corresponds to the range of uncertainties in numerical simulations of topological defects considered in~\cite{Kawasaki:2014sqa}.
}
\label{tab:axion_DM_parameters}
\end{table}

The explicit symmetry breaking term shifts the minimum of the axion effective potential,
\begin{equation}
\theta_{\rm min} \sim  \frac{1}{{\frac{\cal N}{N_{\rm DW}}} + \frac{\chi}{2|g|{\frac{\cal N}{N_{\rm DW}}}M_{\rm Pl}^{4} \(\frac{v_{\rm PQ}}{\sqrt{2}M_{\rm Pl}}\)^{\cal N}}} \Delta \quad \quad \quad ({\rm for\,\,} {\Delta}\ll 1) .
\end{equation}
In order for this shift not to spoil the axion solution to the strong CP problem~\cite{Georgi:1981pu,Ghigna:1992iv,Barr:1992qq,Kamionkowski:1992mf,Holman:1992us,Dine:1992vx},
we impose a constraint on the phase $\Delta$, by forcing the neutron electric dipole moment (nEDM)~\cite{Guo:2015tla},
\begin{equation}
|d_n| \sim 4\times 10^{-16} \times |\theta| \,\,e {\rm \, cm},
\end{equation}
to remain below the experimental limit $d_n<2.9 \times 10^{-26}  \,\,e {\rm \, cm}$~\cite{Baker:2006ts}, implying
$ |\theta_{\rm min}| < 7\times 10^{-11}$.
The constraints on $\Delta$ are shown in table~\ref{tab:axion_DM_parameters}.\footnote{The nEDM constraint can also be avoided if $\Delta$ takes a value close to $\Delta = \pi$.
The degrees of tuning are similar to those shown in table~\ref{tab:axion_DM_parameters}.}
The resulting tuning in the phase $\Delta$ is relatively small compared with the one required to solve the strong CP problem.

The dependence of the upper limit on $\Delta$ on the domain wall number can be understood as follows.
Basically the nEDM constraint can be avoided if $f_a$ is sufficiently small (i.e. the explicit symmetry breaking term is suppressed), except for the cases with
$\Delta \to \pi/2$ for $N_{\rm DW} = 2$, 6 and that with $\Delta \to 3\pi/4$ for $N_{\rm DW} = 4$, where two vacua are degenerate and domain walls never collapse. In those cases the dark matter abundance blows up and a larger value of $|g|$ is required in order to compensate this effect. This large $|g|$ enhances the magnitude of the explicit symmetry breaking term, which violates the nEDM constraint. Therefore, the upper limit on $\Delta$ for benchmarks with HB, RGB, WD appears around $\Delta \simeq \pi/2$ for $N_{\rm DW} = 2$, 6 and $\Delta \simeq 3\pi/4$ for $N_{\rm DW} = 4$.
On the other hand, if $f_a$ is sufficiently large, the nEDM constraint is relevant and some tuning of $\Delta$ is required. The constraint is determined from the parameter dependence
$\theta_{\rm min} \sim |g| N_{\rm DW}^{{\cal N}-1} f_a^{\cal N} \Delta$.
A large factor $N_{\rm DW}^{{\cal N}-1} f_a^{\cal N}$ is compensated by a small factor of $|g|$, which results in the mild $N_{\rm DW}$ dependence for benchmarks with HB, RGB, WD, NS, SN.

The axion features CP violating couplings proportional to $\theta_{\rm min}$, like a Yukawa coupling to nucleons that will be exploited by the ARIADNE experiment to search for axion-mediated long-range forces. Of course, values of $\Delta$ close to the limits quoted in table \ref{tab:axion_DM_parameters} correspond to the maximum CP violation induced by our Planck-suppressed operator and thus the largest and most optimistic effects in ARIADNE, see appendix \ref{app:Sensitivity}.

\section{Discussion and Conclusions}
\label{sec:conclusions}

\begin{figure}[t]
	\centering
	\hspace{-2cm}	\includegraphics[width=0.7\textwidth]{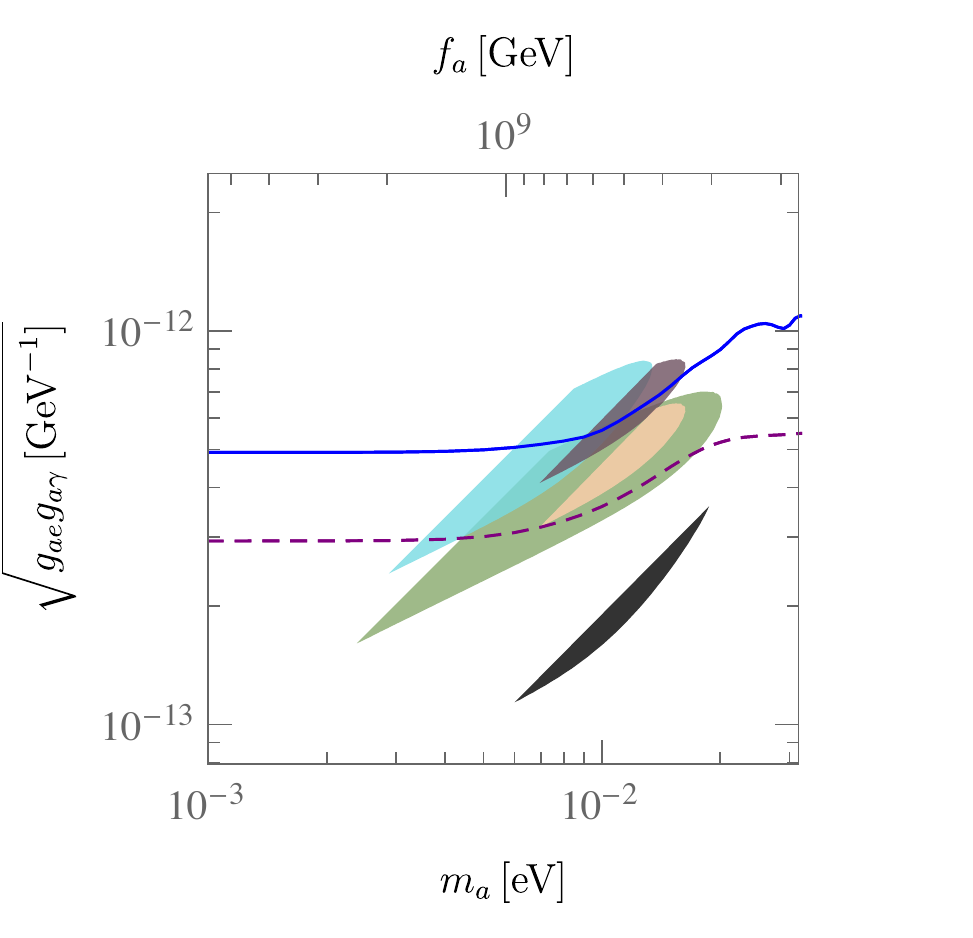}
	\hspace{-1.7cm}\llap{\raisebox{1.5cm}{
			\includegraphics[height=7.1cm]{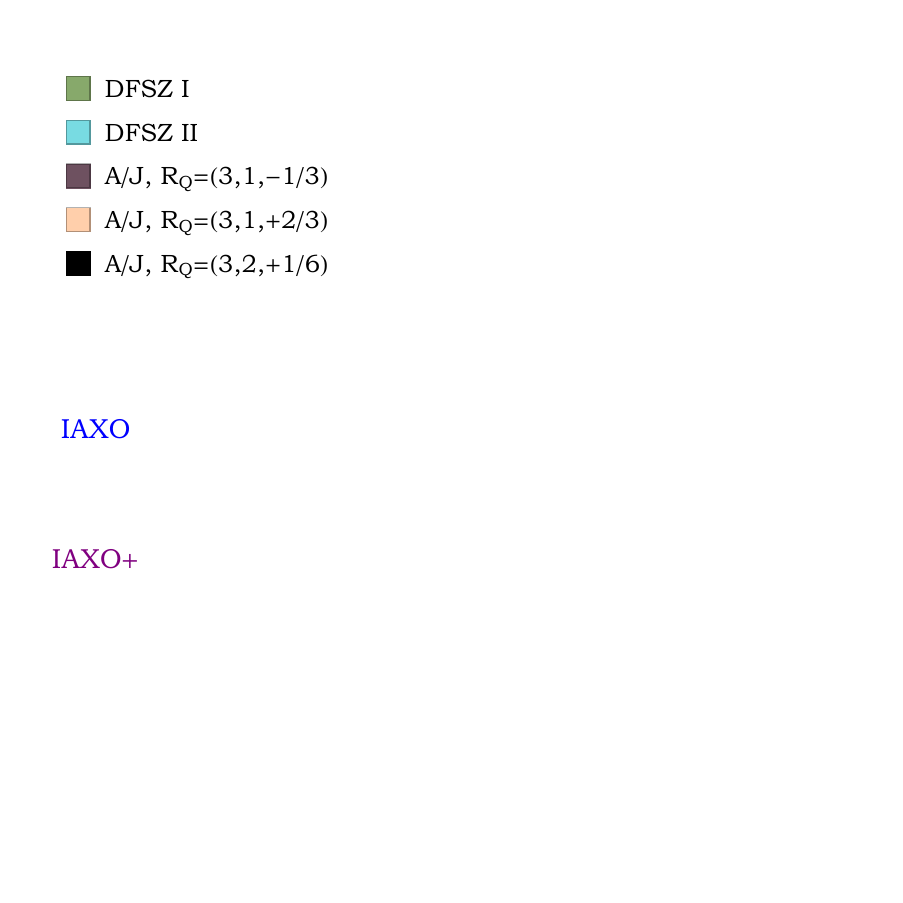}%
	}}
	\caption{2\,$\sigma$ ranges in axion coupling to electrons times coupling to photons vs. axion mass favored by the axion interpretation of the stellar cooling anomalies in WDs, RGBs and HBs, taking into account the constraints from SN 1987A and perturbative unitarity. In the $A/J$ models, also the hint on anomalous cooling from the NS in CAS A is included in the fit. Also shown are the projected sensitivities of IAXO (solid blue) and IAXO+ (dashed purple).
	}
	\label{fig:iaxo_vs_models}
\end{figure}

We have explored quantitatively the possibility that the hints of excessive energy losses of stars in various stages of their
evolution (red giants, helium burning stars, white dwarfs, neutron stars)
can be explained by the axion -- the pseudo Nambu-Goldstone boson predicted by the Peccei-Quinn solution of the
strong CP problem. The main finding of our work is that the overall $>3\,\sigma$ tension of all RG+HBS+WD data taken collectively (and presumably higher if including the NS hint) is successfully removed by taking into account the presence of axions. More specifically, good fits to the data have been obtained in two well-motivated classes of axion models:
DFSZ-type models and KSVZ-type axion/majoron models\footnote{It would be interesting to investigate whether other classes of models, such as the well-motivated axion/flavon models~\cite{Ema:2016ops,Calibbi:2016hwq}, could also produce good fits.}. These fits generically prefer an axion mass around 10 meV,
 if the energy loss constraint from the neutrino pulse duration of SN 1987A is taken into account, cf. figures \ref{fig:hinted_region_DFSZ_tanbeta_m} and \ref{fig:hinted_region_KSVZ_AJ_all}, and tables
\ref{tab:DFSZ_hints}  and \ref{tab:SMASH_fits_wd_rg_hb_plus}.
The predicted regions, compatible with perturbative unitarity at 2\,$\sigma$,  are given in
table \ref{tab:summary_2sigma_ranges} and figure   \ref{fig:iaxo_vs_models}.

\begin{table}[h]
\begin{align}
{\footnotesize
\begin{array}{l | c | c | c |  c | c}
\textbf{Model} 	& m_a [{\rm meV}] 	&  g_{ae} [10^{-13}] & g_{a\gamma} [10^{-12} {\rm GeV^{-1}}]
															& g_{an} [10^{-10}] 	& g_{ap} [10^{-10}] \\
\hline
\vspace{2pt}
{\rm DFSZ\, I}	& 2.4\sim20		&  0.70\sim2.2		&  0.36\sim3.1		& -2.6\sim3.1		& -9.8\sim-2.4	\\
\vspace{2pt}
{\rm DFSZ\, II}	& 2.7\sim 13		&  0.80\sim2.4		&  -3.4\sim-0.7		& -2.1\sim1.9		& -9.4\sim-0.8	\\
\vspace{2pt}
A/J\,\,(3,1,-\frac{1}{3})
			& 7.0\sim 16		&  0.95\sim1.9		&  -4.2\sim-1.8		& -4.9\sim-2.4		& -8.9\sim -3.1	\\ \vspace{2pt}
A/J\,\,(3,1,+\frac{2}{3})
			& 7.0\sim 16		&  0.95\sim1.9		&  1.1\sim2.5		& -4.9\sim-2.4		& -8.9\sim -3.1	\\ \vspace{2pt}
A/J\,\,(3,2,+\frac{1}{6})
			& 6.0\sim 19		&  0.42\sim1.3		&  -1.0\sim-0.30	& -3.6\sim-1.2		& -11\sim -3.6	
\end{array}
}
\nonumber
\end{align}
\caption{
2\,$\sigma$ ranges in masses and couplings to electrons, photons, neutrons, and protons, favored by the axion interpretation of the stellar cooling anomalies in WDs, RGBs and HBs, taking into account the constraints from SN 1987A and perturbative unitarity. In the $A/J$ models, also the hint on anomalous cooling from the NS in CAS A is included in the fit.
}
\label{tab:summary_2sigma_ranges}
\end{table}

Both DFSZ models allow a good fit to the data but DFSZ I has larger proton couplings and thus more tension with the SN 1987A constraints.
None of them gives a good fit to the NS cooling hint so we have not considered it in the 2\,$\sigma$ values given in
table \ref{tab:summary_2sigma_ranges}.
The photon coupling of DFSZ II is larger, making it more suitable for detection in IAXO, cf. figure   \ref{fig:iaxo_vs_models}. The mass range in DFSZ II is narrower.
Axion/majoron models also give a good fit to the data, including this time the NS cooling hint that we have considered in the values given in the table. Masses tend to be larger than in DFSZ and photon couplings as well, with exception of $(3,2,+\frac{1}{6})$ whose small value of $C_{a\gamma}$ makes it quite unfavourable to detect. The reason is that we have excluded values above the perturbative limit of tr$\kappa$, which cuts the large values of $f_a$. The regions are all very close to the perturbative unitarity limit, so they must be taken with a grain of salt.
For these models to give dark matter one needs to include copies of the heavy quarks. In general, this narrows the predicted ranges due to the tension with the perturbativity constraint.

Importantly, the preferred regions in the parameter space can be probed by experiments. 
In particular, the nominal IAXO projection will already be sufficient to probe some of the region of interest while the upgraded IAXO scenario (IAXO+) would decisively test both the DFSZ and the KSVZ $A/J$
interpretation of the stellar energy loss anomalies, as evident from figures \ref{fig:hinted_region_DFSZ_tanbeta_m},  \ref{fig:hinted_region_KSVZ_AJ_all}, and \ref{fig:iaxo_vs_models}. 
In light of this result, the experimental effort to reach the IAXO+ parameters appears highly motivated. 

ARIADNE, on the other hand, shows a capability to access a region of parameter space highly complementary to that of IAXO, which however barely touches the region of interest for DFSZ and does not reach it for KSVZ $A/J$. 

The future ultimate WIMP dark matter experiment DARWIN has a projected sensitivity of $g_{ae}\gtrsim 10^{-12}$ to solar axions \cite{Aalbers:2016jon} and therefore misses the required sensitivity to probe the parameter region of interest from stellar cooling by about one order of magnitude. On the other hand, neutrino detectors such as IceCube, Super-Kamiokande or a future mega-ton water cerenkov detector will probe exactly the parameter region of interest by measuring the neutrino pulse duration of the next galactic SN \cite{Fischer:2016cyd}.
Another possibility to test these axions observationally could be spectral signatures of photon-axion oscillations from celestial objects  \cite{Chelouche:2008ta}.
Moreover, for the DFSZ models, the preferred region of $\tan\beta$ can be complementarily probed by experiments at
the intensity and energy frontier, by searching for direct production of the additional Higgs bosons or for non-SM contributions in flavor physics, cf. e.g. refs.  \cite{Hajer:2015gka,Craig:2016ygr,Arbey:2017gmh}.

If one ignores the SN 1987A constraint on the axion-nucleon couplings, the best fit regions move to slightly higher masses. In these cases (first rows of figures \ref{fig:hinted_region_DFSZ_tanbeta_m} and  \ref{fig:hinted_region_KSVZ_AJ_all}) the relevant region (with the exception of mentioned $R_Q=(3,2,+\tfrac{1}{6})$ models) will be almost entirely probed by IAXO, notably thanks to the buffer gas phase (see appendix~\ref{app:Sensitivity}).

An axion in this mass range can also be the main constituent of dark matter, in particular if the Peccei-Quinn symmetry
is restored after inflation.  Assuming that the PQ symmetry is broken explicitely by a Planck-suppressed operator
${\mathcal L}\supset g M_{\rm Pl}^4 (\sigma/M_{\rm Pl})^9$, the observed dark matter abundance can be
obtained for both DFSZ axions as well as KSVZ axion/majorons with a mass in the range to explain the stellar
cooling anomalies\footnote{We may even consider to extend the DFSZ axion to a DFSZ axion/majoron  \cite{Clarke:2015bea}. Exploiting the radial part of the PQ scalar field as inflaton non-minimally coupled to gravity, the DFSZ $A/J$ and the KSVZ  
$A/J$ models can be considered then as a non-minimal variant of SMASH \cite{Ballesteros:2016euj,Ballesteros:2016xej} -- the acronym standing for SM - Axion - Seesaw - Higgs portal inflation -- explaining six fundamental problems in one stroke: inflation, baryon asymmetry, dark matter, neutrino masses, strong CP problem, stellar cooling anomalies.}, requiring relatively small tuning in $g$, cf. table \ref{tab:axion_DM_parameters}.
However, its direct detection will be a challenge. In this connection it is of interest that in the mass range of
interest, IAXO may indeed measure the mass of the axion \cite{axionmass}. This information could help to design a dedicated
direct detection axion dark matter experiment, for example based on the idea of a dish antenna \cite{Horns:2012jf},
exploiting the axion-photon coupling, such as BRASS. Alternatively, direct axion dark matter experiments
employing the electron coupling, e.g. searches for induced atomic transitions  \cite{Sikivie:2014lha,Braggio:2017oyt},
searches for electron spin precession in the galactic axion dark matter wind, such as QUAX \cite{Barbieri:2016vwg},  and
searches for axion dark matter absorption on a conduction electron, followed by emission of an athermal phonon, in a superconductor
 \cite{Hochberg:2016ajh}, may also ultimately reach the required sensitivity.

\section*{Acknowledgments}

We would like to thank A. Arvanitaki, A. Geraci, J. Hajer, N. Mahmoudi,  and J. Blanchard for discussions and important information.
J.R.\ is supported by the Ramon y Cajal Fellowship 2012-10597 and FPA2015-65745-P
(MINECO/FEDER) as well as the DFG SFB 1258 (Collaborative Research
Center ``Neutrinos, Dark Matter, Messengers'').

\appendix

\section{Data for the global fits}
\label{app:global}

In this section we describe the way we have built our $\chi^2$ model.
For the WD luminosity function, we take the binned LF data points from~\cite{Bertolami:2014noa}, used in \cite{Bertolami:2014wua}, and digitised the synthetic WDLFs as a function of the $g_{ae}$ coupling (the authors present them as function of
$m_a$ with an implicit relation $g_{ae}=0.28\times10^{-13} \, m_a[{\rm meV}] $). We choose 11 points in the range $7<M_{\rm bol}<12.25$.
Higher luminosity points have large error bars, do not help the constraint and decrease unnecessarily the $\chi^2_{\rm min}/$d.o.f.
We take the luminosity as a free-parameter, marginalising over it by finding the smallest $\chi^2$ as a function of $g_{ae}$.
As in \cite{Bertolami:2014wua}, we find a small positive correlation between the normalisation factor and $g_{ae}$. However, it cannot be used to lower the significance of the hint below 3\,$\sigma$. The normalisation is typically a few percent.

We consider 4 WD variables \cite{Corsico:2012sh,Corsico:2014mpa,Corsico:2016okh,Battich:2016htm}.
We exclude G117-B15A~\cite{Corsico:2012ki} from the fit because
it is very similar to R548 (the stars are almost identical, the pulsating mode considered is the same and in both cases there is an issue with the assumption of trapping), which has much more conservative errors.
For each of them we take the measured period decrease with its quoted 1 $\sigma$ error and fit the provided models (again as a function of $g_{ae}$) with their reported  1 $\sigma$ errors. We add each hint as a term in the $\chi^2$ with the 1 $\sigma$ errors in quadrature.

We follow the same procedure for the tip of the RGB of the globular cluster M5 from \cite{Viaux:2013hca,Viaux:2013lha,Arceo-Diaz:2015pva}.
In the case of the $R$-parameter, whose observational value is $R=1.39\pm0.03$, we  employ the theoretical model derived in \cite{Giannotti:2015kwo} as a function of $g_{a\gamma}$ and $g_{ae}$. We take the primordial Helium abundance $Y=0.255\pm0.002$ \cite{Izotov:2014fga} -- which propagates an additional uncertainty in $R$, $\sigma_Y=0.015$ --
and the data from the analysis of~\cite{Ayala:2014pea,Straniero:2015nvc}.

The $\chi^2$ function is then
\begin{eqnarray}
\nonumber
\chi^2 &=& \sum^{\rm WDLF}_{i=1,11}\frac{\(M_i- N(g_{ae}) M(g_{ae})\)^2}{\sigma_{M_i}^2}
+\sum^{\rm WD-var}_{s=1,4}\frac{\(\dot\Pi_s- \dot\Pi_s(g_{ae})\)^2}{\sigma_{\dot\Pi_s}^2+\sigma^2_{\dot\Pi}} \\
&& + \frac{(1.39-R(g_{ae},g_{a\gamma}))^2}{\sigma^2_R+\sigma^2_Y}
+ \frac{(-4.17-M_{\rm TRB}(g_{ae}))^2}{\sigma_{M_{\rm TRB}}^2}.
\end{eqnarray}

For the SN constraint and NS hint we add
\begin{equation}
\(\frac{g_{an}^2+g_{ap}^2}{3.6\times 10^{-19}}\)^2 + \(\frac{g_{an}^2-1.4 \times 10^{-19}}{0.5 \times 10^{-19}}\)^2 .
\end{equation}

\section{Bound on nucleon couplings from SN 1987A}
\label{app:sn1987a}

The observed neutrino signal from SN 1987A has been used to set stringent constraints on the axion-nucleon coupling (see, e.g., \cite{Raffelt:2006cw} for a review).
However, the bounds are based on very few observational data~\cite{Loredo:2001rx,Pagliaroli:2008ur}.
Moreover, the axion production mechanism in the SN is difficult to describe in a complete way and several simplifying (and, often, unjustified) assumptions need to be made~\cite{Fischer:2016cyd}.
Therefore, even the accepted bound from the SN 1987A argument~\cite{Olive:2016xmw},
\begin{align}\label{Eq:SN87A_bound}
 g_{aN} ^2\lesssim 3.57\times 10^{-19}\,,
\end{align}
where  $g_{aN}^2 \equiv m_N^2 C_{aN}^2/f_a^2$ parameterizes the axion-nucleon interaction, has to be taken more as an indicative result than as a sharp limit on the axion-nucleon coupling~\cite{Fischer:2016cyd}.

In this work, we have included the SN 1987A result as an additional constraint for our fits.
More specifically, as discussed at the end of sec.~\ref{app:global}, we have assumed that the axion has a zero coupling to nucleons, with a (1\,$ \sigma $) uncertainty corresponding to
$g_{aN}^2\lesssim 3.6\times 10^{-19}$.
Though this procedure is far from rigorous, it does quantify the effect of the SN 1897A on the hinted parameter space, and it puts it on the same level as the other stellar evolutionary considerations.

A further complication is the choice of the most appropriate parametrization of $ g_{aN} $ in terms of $ g_{an} $ and $ g_{ap} $.
In fact, the axion emission rate, $\varepsilon$, is a function, in general, of both couplings to protons and neutrons.
More specifically, one can show that for nondegenerate nucleons~\cite{Keil:1996ju,Fischer:2016cyd}
\begin{align}\label{Eq:SNepsilon}
\varepsilon\propto \frac{2}{3} \left(I_{\text{nn}} g_{{an}}^2+I_{\text{pp}} g_{{ap}}^2\right)+\frac{2}{9} I_{\text{np}}
\left(g_{{an}}+g_{{ap}}\right)^2+\frac{26}{9} I_{\text{np}} \left(g_{{an}}^2+g_{{ap}}^2\right)\,,
\end{align}
where the functions $ I_{ij} $, corresponding to $ I ({y_i,y_j}) $ in the notation of~\cite{Keil:1996ju}, depend on the stellar plasma conditions.

In the case of the KSVZ axion, the coupling to neutrons is compatible with zero (within the uncertainties).
Therefore eq.~\eqref{Eq:SN87A_bound} leads directly to a bound on the axion-proton coupling~\cite{Olive:2016xmw}.

More general models, however, such as the DFSZ or KSVZ $A/J$, predict interactions to both protons and neutrons, with coupling constants depending on the value of some additional parameter, e.g. $ \tan \beta $ in the case of the DFSZ axion model.
In order to extract the dependence of the constraint~\eqref{Eq:SN87A_bound} on this additional parameter and to use this result in the global fits, we studied numerically the axion rate dependence on the couplings, using the same SN models exploited in~\cite{Fischer:2016cyd}.
The analysis reveals that the term proportional to $ \left(g_{{an}}^2+g_{{ap}}^2\right) $ in Eq.~\eqref{Eq:SNepsilon} is always dominant in the region where the axion emission rate is peaked ($ t\simeq 1 $\,s and $ R\simeq 10 $\,km).
Therefore we adopted the choice
\begin{align}
C_N=\sqrt{C_{ap}^2+C_{an}^2}\,,
\end{align}
corresponding to the constraint in Eq.~\eqref{nu_pulse_sn1987a}.

\section{Projected sensitivity of next generation axion experiments}
\label{app:Sensitivity}

The region of the axion parameter space hinted by the cooling anomalies could be largely accessible to the next generation of axion experiments.
In particular, as discussed in the text, IAXO and ARIADNE show a remarkable potential for both DFSZ and hadronic $A/J$ models.
In this section we discuss our methodology to extract the projected potential of these two experiments to probe the axion parameter space.

\subsection{The IAXO helioscope}

The helioscope is a particularly interesting experimental setup that has already shown the capability to effectively probe axion parameters close to those required to explain the anomalous cooling of stars~\cite{Anastassopoulos:2017ftl}.
Solar axions are transformed into x-ray photons in a magnetic field and then observed in a low background x-ray detector.

Axions can be produced in the Sun through mechanisms involving photons (most notably, the Primakoff process) and electrons (for example, electron bremsstrahlung and Compton processes).
Therefore, the axion production rate depends, in general, on both $ g_{a\gamma} $ and  $ g_{ae} $.
The detection, on the other hand, relies solely on the axion-photon oscillation in a magnetic field and is proportional to $g_{a\gamma}^2 $.

IAXO is a proposal for a next generation axion helioscope~\cite{Irastorza:2011gs}, which will improve of the CAST experiment at CERN, the current state-of-the-art axion helioscope. IAXO would implement a large multibore superconducting magnet with extensive use of x-ray focusing and low-background detection over a large cross-sectional area~\cite{Armengaud:2014gea}, improving current CAST sensitivity by a factor of more than $10^4$ in signal-to-noise ratio.
In this paper we refer to the most updated sensitivity projections of the experiment, as recently produced by the collaboration~\cite{IAXO_potential}.
The information consists of two data sets. The first, $ d_1 (m_a) $, gives the minimal $ \gagamma $ that IAXO can detect if the interaction with electrons is suppressed.
The second, $ d_2 (m_a) $, provides the minimal $ \sqrt{\gagamma g_{ae}} $ that IAXO can detect if axions are produced solely through interaction with electrons. Strictly speaking, these values are the median of the 90\% C.L. excluded values over an ensemble of background-only Monte Carlo simulations. In both cases, we assume $\gagamma$ to be in units of GeV$^{-1}$.

\begin{figure}[ttb]
\centering
\hspace{-0.6cm}
\includegraphics[width=6cm]{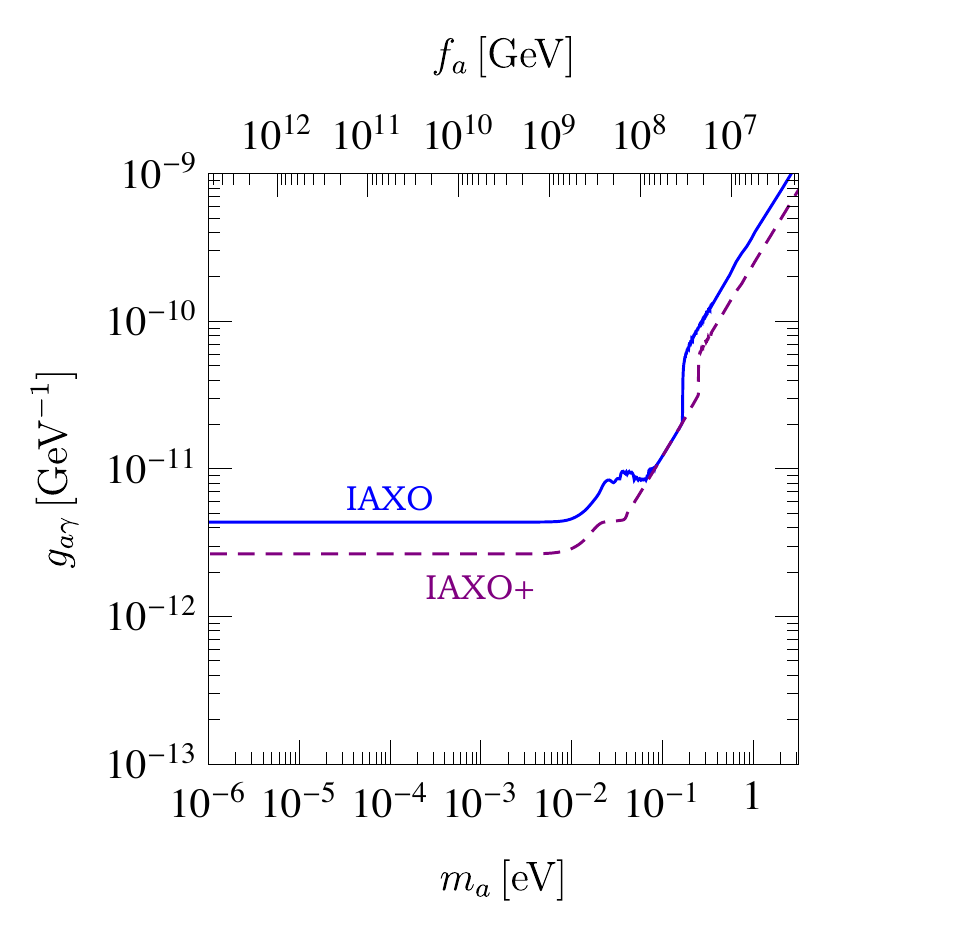}
\hspace{-1cm}
\includegraphics[width=6cm]{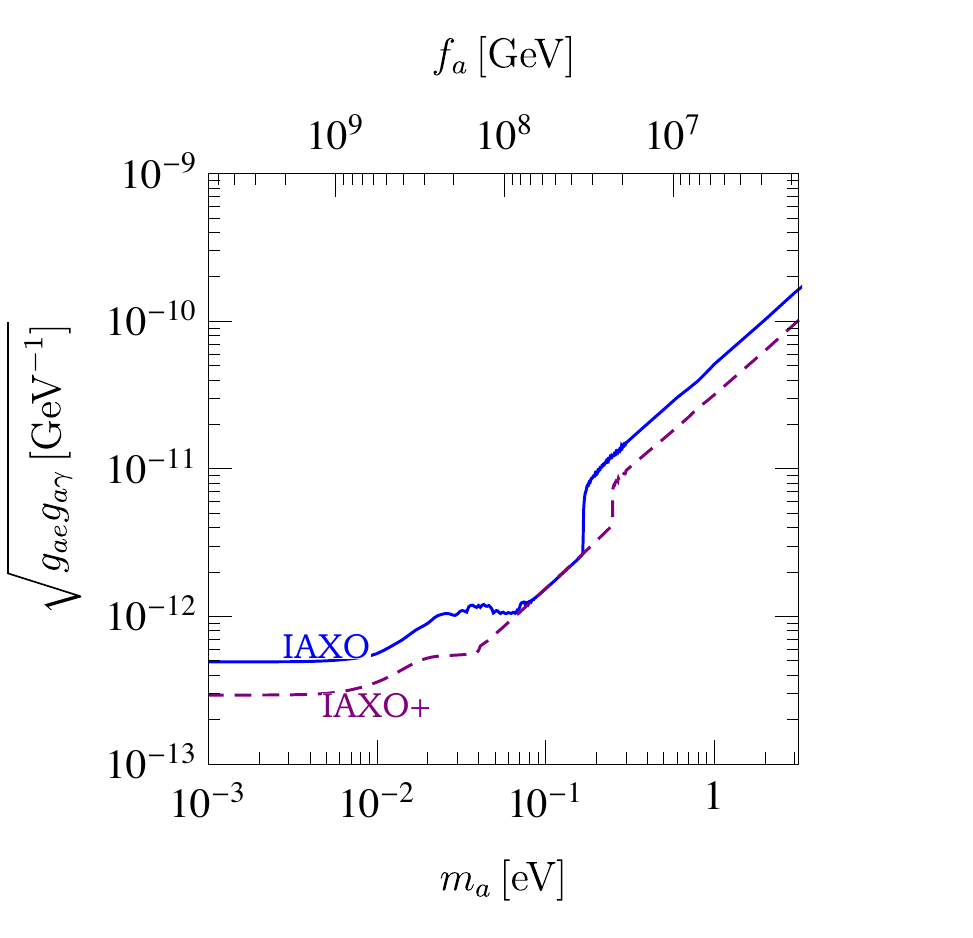}
\hspace{-1.4cm}
\includegraphics[width=6cm]{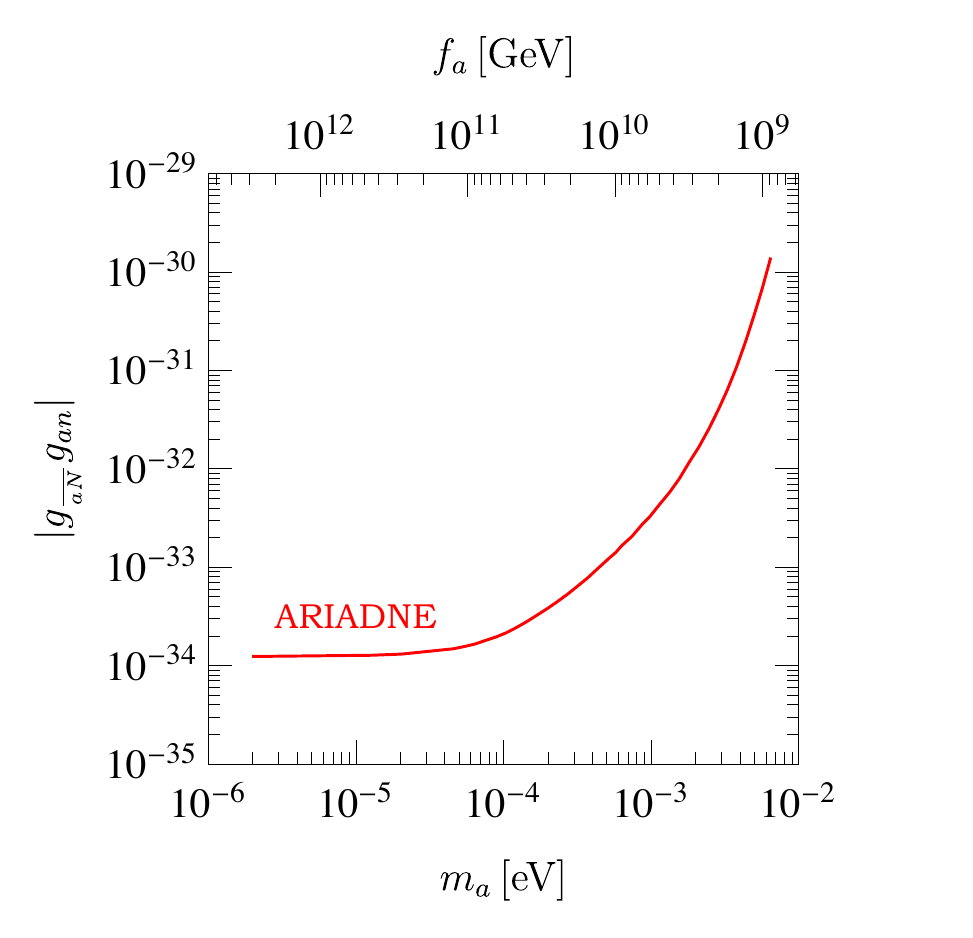}
\hspace{-1cm}

\vspace{-.5cm}
\caption{Projected potential of IAXO and of ARIADNE.
The left and center panels show the data sets $ d_1 $ and $ d_2 $ discussed in the text. The step-wise enhancement at high masses correspond to the planned buffer gas phase~\cite{Armengaud:2014gea}. The right panel shows the ARIADNE sensitivity for the conditions discussed in the text. }
\label{fig:EXP_potential}
\end{figure}

The nominal projection of the experiment following its baseline configuration~\cite{Armengaud:2014gea} is labeled as ``IAXO'' in our plots. The
collaboration has also provided a more optimistic IAXO projection that is based on a series of possible upgraded experimental parameters in both magnet and detectors~\cite{IAXO_potential} beyond their baseline configuration of~\cite{Armengaud:2014gea}. These upgrades would allow for an additional factor $\sim$10 in the signal-to-noise ratio over the nominal projection. This scenario is labeled as ``IAXO+'' in our plots. In view of our results here, the effort of going to this upgraded scenario seems very motivated. The data sets $ d_1 $ and $ d_2 $ are shown in figure~\ref{fig:EXP_potential}, where we use blue for the nominal and purple for the upgraded case.

In general, axions can be produced through interactions with both electrons and photons and so the experimental potential is generally a combination of both datasets, $ d_1 $ and $ d_2 $.
To see this, let us write the axion production rate as
\begin{align}
R=(g_{a\gamma}\times {\rm GeV})^2 P +g_{ae}^2 Q\,,
\end{align}
so that $ P $ and $ Q $ have the same units, and let us indicate with $ (g_{a\gamma}\times {\rm GeV})^2 C $ the conversion probability in the IAXO magnet.
Then, if $ r $ is the distance to the Sun, the minimal number of events expected in IAXO is
\begin{align}\label{Eq:Nmin}
N_{\rm min}= \dfrac{d_1^4 P C}{4\pi r^2}\qquad {\rm or}\qquad N_{\rm min}= \dfrac{d_2^4 Q C}{4\pi r^2}\,,
\end{align}
for the case of axions produced only through interaction with photons ($ Q=0 $) or only through interaction with electrons ($ P=0 $) respectively.
Hence, $ d_1^4 P=d_2^4 Q $.
Notice that this relation does not depend on the axion properties since the axion interaction has been taken out.

For a specific model, one expects $ g_{ae}=\zeta \gagamma $, where we consider $ \gagamma $ in units of GeV$ ^{-1} $, so that $ \zeta $ is dimensionless.
In particular, $ \zeta\simeq 0.20 \sin^2\beta $ for the DFSZ~I model, and $ \zeta\simeq 0.12 \cos^2\beta $ for DFSZ~II.
Hence, in general,
$ R= g_{a\gamma}^{2} P(1+(d_1/d_2)^4\zeta^2)$ and the expected number of events is
\begin{align}\label{Eq:Nevents}
N_{\rm events}=\dfrac{g_{a\gamma}^{4} P  C}{4\pi r^2}
\left( 1+\left( \dfrac{d_1}{d_2}\right)^4 \zeta^2 \right)\,.
\end{align}
Comparing \eqref{Eq:Nevents} with the first relation in \eqref{Eq:Nmin}, we find
\begin{align}\label{Eq:gagmin}
g_{a\gamma}\geq g_{a\gamma}^{\rm min}=d_1 \left( 1+\left( \dfrac{d_1}{d_2}\right)^4 \zeta^2 \right)^{-1/4}\,,
\end{align}
which defines the IAXO potential for a generic axion model.
Notice that $ g_{a\gamma}^{\rm min}\leq d_1 $.
Also, as expected, $ g_{a\gamma}^{\rm min}\simeq d_1 $ for small $ \zeta $, while $ g_{a\gamma}^{\rm min}\simeq d_2/\sqrt{\zeta} $ in the opposite limit.

For the particular case of DFSZ axions, we find
\begin{align}
g_{a\gamma} & = 1.52\times 10^{-10}m_{a} \,, \qquad {\rm DFSZ~I}\,,\\
& = 2.55\times 10^{-10}m_{a} \,, \qquad {\rm DFSZ~II}\,,
\end{align}
where $ m_a $ is in eV and $ \gagamma $ in GeV$ ^{-1} $.
Comparing this to $ g_{a\gamma}^{\rm min} $ in eq.~\eqref{Eq:gagmin} allows to study the mass interval accessible to IAXO for each $ \tan \beta $ (which enters only through $ \zeta $), as shown in figure~\ref{fig:hinted_region_DFSZ_tanbeta_m}.

Notice that the accessible mass interval tends to enlarge monotonically with $ \tan \beta $.
This is easy to understand from eq.~\eqref{Eq:gagmin}, remembering that $ \tan \beta $ enters only through $ \zeta $.
Eventually, for extreme values of $ \tan \beta $, low in the case of DFSZ I and high in the case of DFSZ II, the accessible mass interval becomes constant.
This feature, easily noticeable in figure~\ref{fig:hinted_region_DFSZ_tanbeta_m}, is evident from eq.~\eqref{Eq:gagmin}.
In fact, for $ \zeta $ small enough the dependence from $ \tan \beta $ disappears all together.
It is easy to estimate when this happens.
From the IAXO data sets we notice that $ d_2/d_1 $ is always confined between 5 and 9.\footnote{This is true for both the nominal and upgraded data sets.}
Therefore, $ \zeta^2 (d_2/d_1)^4 \ll 1 $ for $ \zeta\lesssim 10^{-2} $, which corresponds to
$ \tan \beta \lesssim 0.2$ for DFSZ I or $ \tan \beta \gtrsim 0.3$ for DFSZ II,
in good agreement with figure~\ref{fig:hinted_region_DFSZ_tanbeta_m}.

The arguments discussed above imply that, in the DFSZ model, there is a minimal mass range accessible to IAXO, no matter the value of $ \tan \beta $.
These mass ranges are shown in table~\ref{tab:minimal_mass_range}.

\begin{table}[h]
	\begin{align}
	\begin{array}{|c||c|}\hline
	{\rm Model} 		& {\rm mass ~range}~ [{\rm eV}]
	\\	\hline \hline
	 {\rm DFSZ~I} & 0.056,~ 0.17	\\
	 {\rm DFSZ~ II} & 0.029,~ 0.25	\\
	 {{A/J,~ } R_Q=(3,1,-\tfrac{1}{3})} & 0.033,~ 0.17	\\
	{{A/J,~ } R_Q=(3,1,+\tfrac{2}{3})}  & 0.056,~ 0.17	\\
	\hline
	\end{array}
	\nonumber
	\end{align}
	\caption{
		Minimal mass ranges always accessible to IAXO, independently of the other parameter in the model considered.
		See text for discussion.
			}
	\label{tab:minimal_mass_range}
\end{table}

The case of the KSVZ-type axion/majoron, discussed in sec.~\ref{sec:KSVZ-type axion/majoron}, may appear somewhat more complicated to analyze since $ \zeta $ has a dependence on $ m_1 $ through the first term on the right hand side of eq.~\eqref{Eq:axion_majoron_Cae}.
However, this term is always negligible in the region of interest for us.
Repeating the arguments discussed above, we find the IAXO sensitivity regions shown in figure~\ref{fig:hinted_region_KSVZ_AJ_all}.

Notice that also in this case there is a minimal mass region always accessible to IAXO (see table~\ref{tab:minimal_mass_range}), no matter the values of the other parameters, except in the case of the model with $R_Q=(3,2,+\tfrac{1}{6})$, which features a very small photon coupling, cf. eq.~\eqref{eq:photon_coupling}, lowering the sensitivity of IAXO.

\subsection{ARIADNE}
\label{sec:Ariadne_sensitivity}
The most compelling challenge for the axion helioscope is pushing the sensitivity to low masses\footnote{It is also difficult to probe the high mass region, since the axion-photon oscillations loose coherence at large masses.	
However, using a buffer gas it is possible to probe masses up to $ \sim 1$ eV~\cite{Anastassopoulos:2017ftl} and QCD axions with higher masses are excluded by cosmological hot dark matter bounds.
Therefore, here we are mostly concerned with the lower mass challenge.}.
Since the signal in the helioscope is proportional to $ m_a^4 $, its potential diminishes rapidly at low mass.
Thus, probing QCD axion models much below $ m_a \sim 10$ meV is prohibited even with the next generation of axion helioscopes.

Interesting alternatives to the axion helioscope paradigm, which are expected to be better suited to probe the meV mass range,
are experiments that measure axion mediated long range forces~\cite{Moody:1984ba}.
In these cases, the mass controls not only the axion couplings, but also the length of the interaction, lower masses corresponding to longer interaction lengths.
Hence these experiments are expected to be particularly effective in an intermediate mass range around $ m_a\sim  $ meV, corresponding to an interaction length $ ~1/m_a\sim  $ a few 0.1 mm.
A particularly compelling example of these experiments is ARIADNE~\cite{Arvanitaki:2014dfa},
which shows potential to explore sections of the DFSZ axion parameter space.

Among those novel long-range forces, axion dipole-dipole interactions are the least model dependent but very difficult to probe in the near future~\cite{Raffelt:2012sp}.
A more tangible possibility is to measure axion mediated monopole-dipole interactions between a nucleus and a fermion.
These forces are expected to exist if axions have a \textit{CP-violating} interaction\footnote{Here we use convention of a bar above to indicate the CP violating interaction. } with nuclei~\cite{Moody:1984ba}
\begin{align}
{\mathcal L}_{CP}= g_{\over{aN}}\,a\, \bar{N}N\,.
\end{align}

The scalar coupling with nuclei, $ g_{\over{aN}} $, is expected to be nonzero in the SM, because of the CP-violating contribution of the weak interactions~\cite{Georgi:1986kr,Moody:1984ba,Pospelov:1997uv}, while experimental limits on the nEDM constrain it from above,  $ g_{\over{aN}}\leq 10^{-21} (f_a/10^{9}\,{\rm GeV})^{-1} $.

The axion dipole interaction with a fermion, $ g_{af} $, is defined by the interaction Lagrangian~\eqref{axion_leff}.
Integration by parts leads to the usual axion-fermion interaction term
\begin{align}
{\mathcal L}_{af} = -i\dfrac{C_{af} m_f}{f_a}\,a\, \bar \psi_f \gamma_5 \psi_f,
\end{align}
which defines the pseudoscalar coupling $ g_{af}=C_{af} m_f /f_a$, where $ m_f $ is the fermion mass.

ARIADNE searches for forces between a rotating cylinder, made of unpolarized material, and a vessel containing hyper-polarized $ ^3 $He gas~\cite{Arvanitaki:2014dfa}.
Since the $ ^3 $He magnetic moment is dominated by the neutron contribution, the experiment would be measuring the monopole-dipole interaction between nucleus and neutrons, proportional to $ |g_{\over{aN}}g_{an}| $.

Proceeding analogously to what we have done for IAXO, we can define the ARIADNE sensitivity  AS($ m_a $)  as the minimal value of $ |g_{\over{aN}}g_{an}| $ that the experiment is expected to measure.
For our analysis, we have extracted the AS from~\cite{Arvanitaki:2014dfa}, assuming the most optimistic value for the scalar coupling
$ g_{\over{aN}}=10^{-21} (f_a/10^{9}\,{\rm GeV})^{-1} $, a total integration time of $ 10^6 $\,s and a transverse relaxation time of $ T_2 =1000 $\,s (cf. figure~2 in~\cite{Arvanitaki:2014dfa}).
The resulting AS is shown in the right panel of figure~\ref{fig:EXP_potential}.

The axion parameter space accessible to ARIADNE is the set of $ m_a $ and $ \tan \beta $ (for the DFSZ model) or $ m_a $ and $ \tr \,\kappa $ (for the A/J models) satisfying
\begin{align}\label{Eq:ARIADNE_criterion}
\log \left( |g_{\over{aN}} g_{an}|/{\rm AS}(m_a)\right) \geq 0\,.
\end{align}
This region is shown in figures~\ref{fig:hinted_region_DFSZ_tanbeta_m} and \ref{fig:hinted_region_KSVZ_AJ_all}.

As evident from figure~\ref{fig:hinted_region_DFSZ_tanbeta_m}, ARIADNE could probe particularly well the low mass end of the DFSZ axion parameter space relevant for the cooling anomalies.
There is therefore an interesting complementarity between IAXO and ARIADNE.
The potential for the A/J models is less impressive.

Notice that, contrarily to IAXO, ARIADNE cannot cover the whole range of $ \tan \beta $ for the DFSZ models.
In fact, the DFSZ axion interaction with neutrons vanishes for $ \tan \beta \simeq 0.8 $, and $ |g_{\over{aN}}g_{an}|  $ becomes inaccessibly small for $ \tan \beta \sim 0.1-1$.
The same is true for the KSVZ $A/J$ models, in which the coupling to neutrons is suppressed at low values of $ \tr \,\kappa $, and obviously for the pure KSVZ axions, in which the coupling to neutrons is compatible with zero.
Therefore, in the case of ARIADNE there is not a minimal mass range accessible for all values of the other parameters, at least for the DFSZ and KSVZ axion models.

It is interesting to point out that these problems are inherent in the choice of the $ ^3 $He gas, in which the proton contribution to the magnetic moment vanishes.
The choice of $ ^3 $He has undiscussed experimental advantages, most notably a long coherence time and the possibility to be highly polarized~\cite{Arvanitaki:2014dfa}.
However, as we have seen, in most realistic models the axion couples to protons more strongly than to neutrons and, more importantly, the neutron interaction vanishes for some values of the parameters in all models considered here.
Therefore, it would be certainly very beneficial to develop an alternative experimental setup which does not rely exclusively on the $ g_{an} $ coupling for detection.

Finally, proceeding as we did above, it is straightforward to show that the future projection reach of the experiment, as discussed in~\cite{Arvanitaki:2014dfa} and shown with a solid blue line in figure~2 therein, would allow to cover most of the DFSZ axion parameter space.
The possibility to scale the experiment to reach such sensitivity is feasible though, most likely, not in the very near future.
Remember, however, that the projected sensitivity is still based on the assumption that the CP-violating scalar coupling is the largest possible allowed by measurements of the nEDM, an assumption not well justifiable.
This requirement can be removed only by measuring dipole-dipole interactions, which do not depend on the scalar coupling.
This would be in principle possible, using a polarized source mass, but it would present considerable more experimental difficulties.
As shown in figure 3 of~\cite{Arvanitaki:2014dfa}, reaching the parameter space relevant to realistic axion models in this case is, at the moment, still prohibitive.

\end{document}